\documentclass[%
 reprint,
superscriptaddress,
pre,
 amsmath,amssymb,
 aps,
]{revtex4-1}

\usepackage{graphicx}
\usepackage{dcolumn}
\usepackage{bm}
\usepackage[colorlinks= true, linkcolor=blue, citecolor=blue, urlcolor=blue]{hyperref}
\usepackage{lipsum}
\usepackage{mathtools}
\usepackage{xcolor}
\usepackage{multirow}

\begin{document}

\title{Fitness dependence of the fixation-time distribution \\ for evolutionary dynamics on graphs}

\author{David Hathcock}
\affiliation{Department of Physics,  Cornell University, Ithaca, New York 14853, USA}
\author{Steven H. Strogatz}
\affiliation{Department of Mathematics,  Cornell University, Ithaca, New York 14853, USA}

\date{\today}

\begin{abstract}
Evolutionary graph theory models the effects of natural selection and random drift on structured populations of competing mutant and non-mutant individuals. Recent studies have found that fixation times in such systems often have right-skewed distributions. Little is known, however, about how these distributions and their skew depend on mutant fitness. Here we calculate the fitness dependence of the fixation-time distribution for the Moran Birth-death process in populations modeled by two extreme networks: the complete graph and the one-dimensional ring lattice, obtaining exact solutions in the limit of large network size. We find that with non-neutral fitness, the Moran process on the ring has normally distributed fixation times, independent of the relative fitness of mutants and non-mutants. In contrast, on the complete graph, the fixation-time distribution is a fitness-weighted convolution of two Gumbel distributions. When fitness is neutral the fixation-time distribution jumps discontinuously and becomes highly skewed on both the complete graph and the ring. Even on these simple networks, the fixation-time distribution exhibits rich fitness dependence, with discontinuities and regions of universality. Extensions of our results to two-fitness Moran models, times to partial fixation, and evolution on random networks are discussed.
\end{abstract}

\pacs{Valid PACS appear here}
\maketitle

\section{Introduction}

Reproducing populations undergo evolutionary dynamics. Mutations can endow individuals with a fitness advantage, allowing them to reproduce more quickly and outcompete non-mutant individuals \cite{nowak2006evolutionary}. Two natural questions arise: If a single mutant individual is introduced into a population, what is the \emph{probability} that the mutant lineage will spread and ultimately take over the population (an outcome known as fixation)? And if  fixation occurs, how much \emph{time} does it take?

These questions have been addressed, in part, by evolutionary graph theory, which studies evolutionary dynamics in structured populations. Thanks to this approach, fixation probabilities are now well understood for various models on various networks \cite{maruyama1974markov, maruyama1974simple, slatkin1981fixation, lieberman2005evolutionary, antal2006fixation, houchmandzadeh2011fixation, diaz2014approximating, kaveh2015duality, jamiesonlane2015fixation, altrock2017evolutionary, tkadlec2019population}. Less is known about fixation times. 
Given a model of evolutionary dynamics, one would like to predict the mean, variance, and ideally the full distribution of its fixation times. 

Of these quantities, the mean is the best understood. Numerical and analytical results exist for mean fixation times on both deterministic \cite{kimura1980average, slatkin1981fixation, antal2006fixation, altrock2009fixation, frean2013effect, askari2015analytical, altrock2017evolutionary, askari2017effect, tkadlec2019population} and random \cite{askari2015analytical, askari2017effect, farhang2017effect, hajihashemi2019fixation} networks. Yet although mean fixation times are important to study, the information they provide can be misleading, because fixation-time distributions tend to be broad and skewed and hence are not well characterized by their means alone \cite{dingli2007stochastic, altrock2011stability, ashcroft2015mean,altrock2017evolutionary, ying2018mean}. Initial analytical results have determined the asymptotic fixation-time distribution for several simple networks, but only when the relative fitness of the mutants  is infinite \cite{aldous2013interacting, ottino2017takeover, ottino2017evolutionary}. For other values of the relative fitness, almost nothing is known. Preliminary results suggest that at neutral fitness (when mutants and non-mutants are equally fit), the fixation-time distribution becomes highly right-skewed \cite{ottino2017evolutionary}. 

In this paper we investigate the full fitness dependence of fixation-time distributions for the Moran process \cite{moran1958random, moran1958effect}, a simple model of evolutionary dynamics. In the limit of large network size, we derive asymptotically exact results for the fixation-time distribution and its skew for two network structures at opposite ends of the connectivity spectrum: the complete graph, in which every individual interacts with every other individual; and the one-dimensional ring lattice, in which each individual interacts only with its nearest neighbors on a ring.
 
The specific model we consider is the Moran Birth-death (Bd) process
    \footnote{We use the convention that capital letters designate a fitness dependent step in the Moran process (e.g. for the Bd process nodes give birth at a rate proportional to their fitness, but die with uniform probability). See Ref.~\cite[Box 2]{ottino2017evolutionary} for a detailed explanation of this nomenclature.}, defined as follows. On each node of the network there is an individual, either mutant or non-mutant. The mutants have a fitness level $r$, which designates their relative reproduction rate compared to non-mutants. When $r>1$, the mutants have a fitness advantage, whereas when $r=1$ they have neutral fitness. At each time step we choose a node at random, with probability proportional to its fitness, and choose one of its neighbors with uniform probability. The first individual gives birth to an offspring of the same type. That offspring replaces the neighbor, which dies. The model population is updated until either the mutant lineage takes over (in which case fixation occurs) or the mutant lineage goes extinct (a case not considered here).

As mentioned above, the distribution of fixation times is often skewed. The skew emerges from the stochastic competition between mutants and non-mutants through multiple mechanisms. For instance, when the mutants have neutral fitness the process resembles an unbiased random walk. We find that the asymptotic fixation-time distribution for a simple random walk is only skewed when the walk is unbiased. The lack of bias allows for occasional long recurrent excursions (that are suppressed in biased walks) during successful runs to fixation. The fixation-time distribution is strongly skewed because there are many ways to execute such walks that are much longer than usual, but comparably few ways for mutants to sweep through the population much faster than usual.

Depending on network structure, the fixation-time skew can also come from a second, completely separate mechanism, which involves characteristic slowdowns that arise because individuals do not discriminate between mutants and non-mutants during the replacement step of the Moran process. For example, when very few non-mutants remain, the mutants can waste time replacing each other. These slowdowns are reminiscent of those seen in a classic problem from probability theory, the coupon collector's problem, which asks: How long does it take to complete a collection of $N$ distinct coupons if a random coupon is received at each time step? The intuition for the long slowdowns is clear: when nearly all the coupons have been collected, it can take an exasperatingly long time to collect the final few, because one keeps acquiring coupons that one already has. The problem was first solved by Erd\H{o}s and R\'{e}nyi, who proved that for large $N$, the time to complete the collection has a Gumbel distribution \cite{erdos1961classical}. In fact, for evolutionary processes with infinite fitness there exists an exact mapping onto coupon collection \cite{ottino2017evolutionary, ottino2017takeover}. Remarkably, while this correspondence breaks down for finite fitness, the coupon collection heuristic still allows us to predict correct asymptotic fixation-time distributions for non-neutral fitness.

In the following sections we show that for $N\gg1$, the neutral-fitness Moran process on the complete graph and the one-dimensional ring lattice has highly skewed fixation-time distributions, and we solve for their cumulants exactly. For non-neutral fitness the fixation-time distribution is normal on the lattice and a weighted convolution of Gumbel distributions on the complete graph. These results are novel; apart from the infinite fitness limit and some partial results at neutral fitness (noted below), the fitness dependence of these distributions was previously unknown.

We begin by developing a general framework for computing fixation-time distributions and cumulants of birth-death Markov chains, and then apply it to the Moran process to prove the results above. We also consider the effects of truncation on the process and examine how long it takes to reach partial, rather than complete, fixation. The fixation-time distributions have rich dependence on the fitness level and the degree of truncation, with both discontinuities and regions of universality. To conclude, we discuss extensions of our results to two-fitness Moran models and to more complicated network topologies.

\section{General Theory for Birth-Death Markov Processes}\label{generalTheory}

For simplicity, we restrict attention to network topologies and initial mutant populations for which the probability of adding or removing a mutant in a given time step depends only on the number of existing mutants, not on where the mutants are located on the network. The state of the system can therefore be defined in terms of the number of mutants, $m = 0, 1, \dots, N$, where $N$ is the total number of nodes on the network. The Moran process is then a birth-death Markov chain with $N+1$ states, transition probabilities $b_m$ and $d_m$ determined by the network structure, and absorbing boundaries at $m=0$ and $m=N$. In this section we review several general analytical results for absorbing birth-death Markov chains, explaining how they apply to fixation times in evolutionary dynamics. We also develop an approach, which we call \emph{visit statistics}, that enables analytical estimation of the asymptotic fixation time cumulants.

On more complicated networks, the probability of adding or removing a mutant depends on the configuration of existing mutants. For some of these networks, however, the transition probabilities can be accurately estimated using a mean-field approximation \cite{hajihashemi2019fixation, ying2018mean, ottino2017evolutionary, ottino2017takeover}. Then, to a good approximation, the results below apply to such networks as well.

\subsection{Eigendecomposition of the birth-death process}\label{eigenDecomp}

Assuming a continuous-time process, the state of the Markov chain described above evolves according to the master equation,
\begin{equation}\label{masterEq}
\dot{\mathbf{p}}(t) = \Omega \cdot \mathbf{p}(t),
\end{equation}
where $\mathbf{p}(t)$ is the probability of occupying each state of the system at time $t$ and $\Omega$ is the transition rate matrix, with columns summing to zero. In terms of the transition probabilities $b_m$ and $d_m$, the entries of $\Omega$ are
\begin{equation}\label{transitionMatrix}
\Omega_{m n} = b_n \delta_{m, n+1}+ d_n \delta_{m, n-1} - (b_n + d_n) \delta_{m,n},
\end{equation}
where $m$ and $n$ run from $0$ to $N$, $\delta_{m,n}$ is the Kronecker delta, and $b_0 = d_0 = b_N = d_N = 0$. The final condition guarantees the system has absorbing boundaries with stationary states $p_m = \delta_{m, 0}$ and $p_m = \delta_{m, N}$ when the population is homogeneous. Thus we can decompose the transition matrix into stationary and transient parts, defining the transient part $\Omega_\text{tr}$ as in Eq.~(\ref{transitionMatrix}), but with $m,\, n = 1, \dots, N-1$. The transient transition matrix acts on the transient states of the system, denoted $\mathbf{p}_\text{tr}(t)$. The eigenvalues of $\Omega_\text{tr}$ are real and strictly negative, since probability flows away from these states toward the absorbing boundaries. To ease notation in the following discussion and later applications, we shall refer to the positive eigenvalues of $-\Omega_\text{tr}$ as the eigenvalues of the transition matrix, denoted $\lambda_m$, where $m = 1, \dots, N-1$.

From the perspective of Markov chains, the fixation time $T$ is the time required for first passage to state $m=N$, given $m_0$ initial mutants, $p_m(0) = \delta_{m,m_0}$. At time $t$, the probability that state $N$ has been reached (i.e., the cumulative distribution function for the first passage times) is simply $\varphi_{m_0}^{-1} p_N(t)$, where $\varphi_{m_0}$ is the fixation probability given $m_0$ initial mutants. The distribution of first passage times is therefore $\varphi_{m_0}^{-1} \dot{p}_N (t) = \varphi_{m_0}^{-1} b_{N-1} p_{N-1}(t)$. Since we normalize by the fixation probability, this is precisely the fixation-time distribution conditioned on reaching $N$.

The solution to the transient master equation is the matrix exponential $\mathbf{p}_\text{tr}(t) = \exp(\Omega_\text{tr} t) \cdot \mathbf{p}_\text{tr}(0)$, yielding a fixation-time distribution $\varphi_{m_0}^{-1}b_{N-1} [\exp(\Omega_\text{tr} t) \cdot \mathbf{p}_\text{tr}(0)]_{N-1}$ \cite{asmussen2003applied}. If we assume one initial mutant $m_0=1$ this becomes $\varphi_1^{-1} b_{N-1} [\exp(\Omega_\text{tr} t)]_{N-1, 1}$. The matrix exponential can be evaluated in terms of the eigenvalues $\lambda_m$ by taking a Fourier (or Laplace) transform (for details, see Ref.~\cite{ashcroft2015mean}). For a single initial mutant, the result is that the fixation time $T$ has a distribution $f_T(t)$ given by
\begin{equation}\label{fixation_distribution}
f_T(t) = \sum_{j=1}^{N-1} \left(\prod_{k = 1, k\neq j}^{N-1} \frac{\lambda_k}{\lambda_k - \lambda_j} \right) \lambda_j e^{-\lambda_j t}.
\end{equation}
This formula holds as long as the eigenvalues $\lambda_m$ are distinct, which for birth-death Markov chains occurs when $b_m$ and $d_m$ are non-zero (except at the absorbing boundaries) \cite{keilson1979markov}. Generalizations of this result for arbitrarily many initial mutants have also recently been derived, in terms of eigenvalues of the transition matrix and certain sub-matrices \cite{ashcroft2015mean}. 

The distribution in Eq.~(\ref{fixation_distribution}) is exactly that corresponding to a sum of exponential random variables with rate parameters $\lambda_m$. The corresponding cumulants equal $(n-1)! \sum_{m = 1}^{N-1} (\lambda_m)^{-n}$. As our primary interest is the asymptotic shape of the distribution, we normalize $T$ to zero mean and unit variance and study $(T-\mu)/\sigma$, where $\mu$ and $\sigma$ denote the mean and standard deviation of $T$. The standardized distribution is then given by $\sigma f_T(\sigma t + \mu)$. The rescaled fixation time has cumulants  
\begin{equation}\label{fixation_cumulants}
\kappa_n(N) = (n-1)! \left( \sum_{m =1}^{N-1} \frac{1}{\lambda_m^n} \right) \Bigg/  \left( \sum_{m =1}^{N-1} \frac{1}{\lambda_m^2} \right)^{n/2},
\end{equation}
which, for many systems including those considered below, are finite as $N \rightarrow \infty$. When the limit exists, we define the asymptotic cumulants by $\kappa_n = \lim_{N\rightarrow\infty} \kappa_n(N)$. In particular, because we have standardized our distribution, the third cumulant $\kappa_3$ is the skew. In practice the limit $N\rightarrow \infty$ is taken by computing the leading asymptotic behavior of both the numerator and denominator in Eq.~(\ref{fixation_cumulants}). As we will see below the scaling of these terms with $N$ depends on both the population network structure and the mutant fitness (see also asymptotic analysis in Supplemental Material, Sections S3 \& S4 \cite{SM}). This approach allows us to characterize the asymptotic shape of the fixation-time distribution in terms of the constants $\kappa_n$. Since $\lambda_m>0$, it is clear from this expression that, for finite $N$, the skew and all higher order cumulants must be positive, in agreement with results for random walks with non-uniform bias \cite{bakhtin2018universal}. As $N\rightarrow \infty$ this is not necessarily true; in some cases the cumulants vanish.

The eigendecomposition gives the fixation-time distribution and cumulants in terms of the non-zero eigenvalues of the transition matrix. In general the eigenvalues must be found numerically, but in cases where they have a closed form expression the asymptotic form of the cumulants and distribution can often be obtained exactly.

\subsection{Analytical cumulant calculation: Visit statistics}\label{visitStatistics}

In this section we develop machinery to compute the cumulants of the fixation time analytically without relying on matrix eigenvalues. For this analysis, we specialize to cases where $b_m/d_m = r$ for all $m$, relevant for the Moran processes considered below. These processes can be thought of as biased random walks overlaid with non-constant waiting times at each state. 

It is helpful to consider the Markov chain conditioned on hitting $N$, with new transition probabilities $\tilde{b}_m$ and $\tilde{d}_m$ so that the fixation probability $\varphi_{m_0} = 1$. If $X_t$ is the state of the system at time $t$, then $\tilde{b}_m = \mathcal{P}(X_t = m \rightarrow X_{t+1} = m+1 | X_\infty = N)$ with $\tilde{d}_m$ defined analogously. We derive explicit expressions for $\tilde{b}_m$ and $\tilde{d}_m$ in Supplemental Material, Section S1 \cite{SM}. Conditioning is equivalent to a similarity transformation on the transient part of the transition matrix: $\tilde{\Omega}_\text{tr} = S \, \Omega_\text{tr} \, S^{-1}$, where $S$ is diagonal with $S_{mm} = 1-1/r^m$. Furthermore, since $b_m/d_m = r$, we can decompose $\Omega_\text{tr} = \Omega_\text{RW} D$, where $D$ is a diagonal matrix, $D_{mm} = b_m + d_m$, that encodes the time spent in each state and $\Omega_\text{RW}$ is the transition matrix for a random walk with uniform bias,
\begin{equation}
    [\Omega_\text{RW}]_{nm} = \frac{r}{1+r} \delta_{m, n+1}+ \frac{1}{1+r} \delta_{m, n-1} -  \delta_{m,n}.
\end{equation}
Applying the results of the previous section and using the fact that the columns of $\Omega$ sum to zero, we can write there fixation-time distribution of the conditioned Markov chain as $f_T(t) = -\mathbf{1}\tilde{\Omega}_\text{tr} \exp(\tilde{\Omega}_\text{tr} t) \mathbf{p}_\text{tr}(0)$, where $\mathbf{1}$ is the row vector containing all ones. This distribution has characteristic function \cite{asmussen2003applied}
\begin{equation}
    \phi(\omega) \coloneqq E[\exp(i\omega T)] = \mathbf{1} \tilde{\Omega}_\text{tr} (i \omega + \tilde{\Omega}_\text{tr})^{-1} \mathbf{p}_\text{tr}(0).
\end{equation}
and the derivatives $(-i)^n \phi^{(n)}(0)$ give the moments of $T$
\begin{equation}\label{fixationMoments}
    E[T^n] = (-1)^n n! \mathbf{1} \tilde{\Omega}_\text{tr}^{-n} \mathbf{p}_\text{tr}(0),
\end{equation}
in terms of $\tilde{\Omega}_\text{tr}^{-1} = D^{-1} S \Omega_\text{RW}^{-1} S^{-1}$. This inverse has a nice analytical form because $S$ and $D$ are diagonal and $\Omega_\text{RW}$ is tridiagonal Toeplitz. We call this approach \emph{visit statistics} because the elements $V_{ij}$ of $V = -S \Omega_\text{RW}^{-1} S^{-1}$ encode the average number of visits to state $i$ starting from state $j$.

Each power of $\tilde{\Omega}_\text{tr}$ in Eq.~(\ref{fixationMoments}) produces products of $(b_i + d_i)$ that arise in linear combinations determined by the visit numbers $V_{ij}$. Therefore, the cumulants of the fixation time have the general form 
\begin{equation}\label{exactCumulants}
    \kappa_n(N) = \frac{\displaystyle \sum_{i_1,i_2, \dots,i_n=1}^{N-1} \frac{w_{i_1i_2 \cdots i_n}^n(r,N | m_0)}{(b_{i_1}+d_{i_1})(b_{i_2}+d_{i_2})\cdots(b_{i_n}+d_{i_n})}}{ \displaystyle \left(\sum_{i,j=1}^{N-1} \frac{w_{ij}^2(r,N |m_0)}{(b_i+d_i)(b_j+d_j)} \right)^{n/2}},
\end{equation}
where $w_{i_1i_2 \cdots i_n}^n(r,N|m_0)$ are weighting factors based on the visit statistics of the biased random walk, given the initial number of mutants $m_0$. In what follows, we always assume $m_0=1$ and suppress the dependence of the weighting factors on initial condition, writing $w_{i_1i_2 \cdots i_n}^n(r,N)$ instead. A detailed derivation of Eq.~(\ref{exactCumulants}) and explicit expressions for $w_{ij}^2(r,N)$ and $w_{ijk}^3(r,N)$ are given in the Appendix below.

To the best of our knowledge this representation of the fixation-time cumulants has not been previously derived, although a similar approach was recently used to compute mean fixation times for evolutionary dynamics on complex networks \cite{hajihashemi2019fixation}. This expression is equivalent to the well-known recurrence relations for absorption-time moments of birth-death processes \cite{altrock2011stability, goel1974stochastic} but is easier to handle asymptotically, and can be useful even without explicit expressions for $w_{i_1i_2 \cdots i_n}^n(r,N)$. Estimating the sums in Eq.~(\ref{exactCumulants}) allows us to compute the asymptotic fixation time cumulants exactly.

\subsection{Recurrence relation for fixation-time moments}\label{sec:reccurence}

Evaluation of the eigenvalues of the transition matrix for large systems can be computationally expensive, with the best algorithms having run times quadratic in matrix size. Numerical evaluation of the expression given in Eq.~(\ref{exactCumulants}) is even worse, as it requires summing $\mathcal{O}(N^n)$ elements. If only a finite number of fixation time cumulants (and not the full distribution) are desired, there are better numerical approaches. Using standard methods from probability theory \cite{keilson1965review}, we derive a recurrence relation that allows numerical moment computation with run time linear in system size $N$. For completeness we provide the full derivation of the reccurence for the fixation-time skew in Supplemental Material, Section S2 \cite{SM}.

\subsection{Equivalence between advantageous and disadvantageous mutations}\label{disadvantageousMutants}
In the following applications, we will generally speak of the mutants as having a fitness advantage, designated by the parameter $r>1$. Our results, however, can be immediately extended to disadvantageous mutations. In particular, the fixation-time distributions (conditioned on fixation occurring) for mutants of fitness $r$ and $1/r$ are identical. When a mutant with fitness $1/r$ is introduced into the population (and eventually reaches fixation), the non-mutants are $r$ times as fit as the mutants. Therefore, this system is equivalent to another system that starts with $N-1$ fitness $r$ mutants which eventually die out (the mutants in the former system are the non-mutants in the latter). It has been shown that the times to go from one initial mutant to fixation ($m=1 \rightarrow m=N$) and from $N-1$ initial mutants to extinction ($m=N-1 \rightarrow m=0$) have identical distribution \cite{ashcroft2015mean}. Thus indeed, the conditioned fixation-time distributions are identical for mutants of fitness $r$ and $1/r$. Of course the fixation probability is very different in the two cases: for the disadvantageous mutations it approaches 0 for large $N$ \cite{lieberman2005evolutionary}.

\section{One-Dimensional Lattice}\label{1D_lattice}

\begin{figure*}[t]
\includegraphics[width=0.49\linewidth]{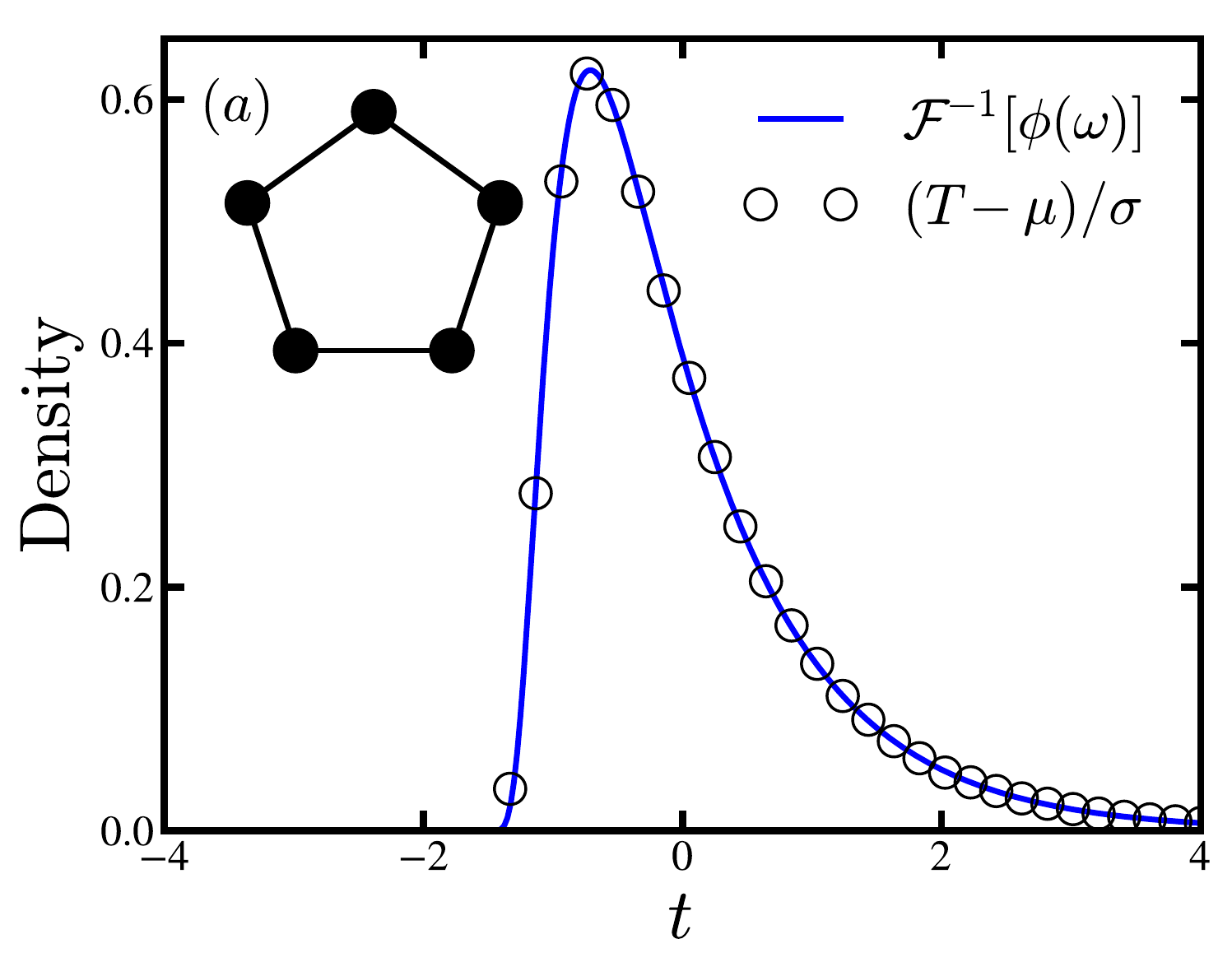}\,\,\,\, \includegraphics[width=0.49\linewidth]{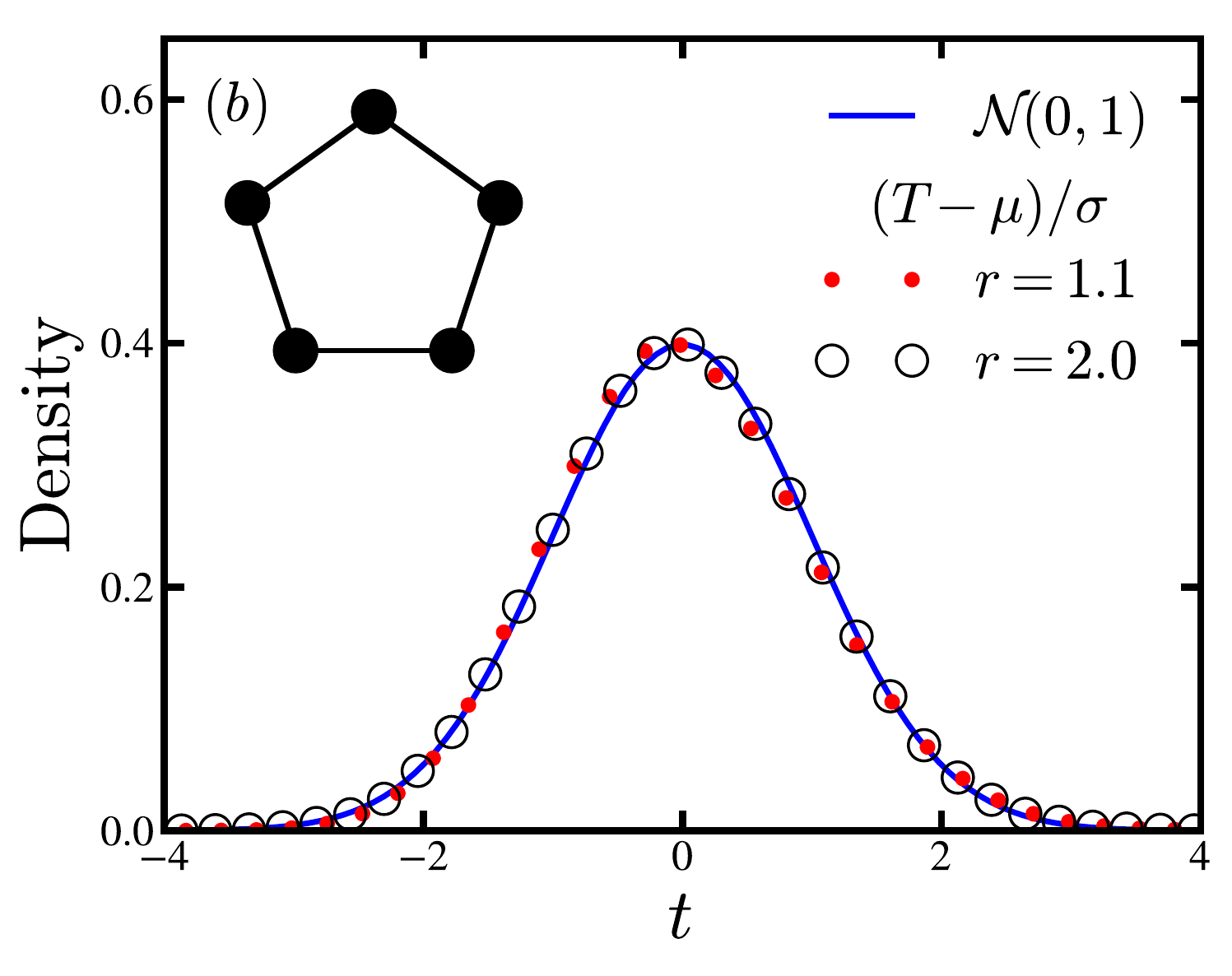}
\caption{\label{fig:ringDistributions} Fixation-time distributions on the 1D lattice obtained from $10^6$ simulation runs. All distributions are standardized to zero mean and unit variance. Solid curves are the theoretical predictions. Shown are the fixation-time distributions for (a) a 1D lattice of $N=100$ nodes with neutral fitness  $r = 1$ and (b) a 1D lattice of $N = 5000$ nodes with mutant fitnesses $r=1.1$ and $r = 2.0$. For the neutral fitness case, the theoretical distribution was generated by numerical inverse Fourier transform of the characteristic function (Eq.~(\ref{1DneutralChar})). The $r=1.1$ distribution is  slightly but visibly skewed due to finite network size.}
\end{figure*}

We now specialize to Moran Birth-death (Bd) processes, starting with the one-dimensional (1D) lattice. We assume periodic boundary conditions, so that the $N$ nodes form a ring. The mutants have relative fitness $r$, meaning they give birth $r$ times faster, on average, than  non-mutants do.

Starting from one mutant, suppose that at some later time $m$ of the $N$ nodes are mutants. On the 1D lattice, the population of mutants always forms a connected arc, with two mutants at the endpoints of the arc. Therefore, the probability $b_m$ of increasing the mutant population by one in the next time step is the probability of choosing a mutant node at an endpoint to give birth, namely $2 r/(rm+N-m)$,  times the probability $1/2$ that the neighboring node to be replaced is not itself a mutant. (The latter probability equals 1/2 because there are two neighbors to choose for replacement: a mutant neighbor on the interior of the arc and a non-mutant neighbor on the exterior. Only the second of these choices produces an increase in the number of mutants.) Multiplying these probabilities together we obtain
\begin{equation}\label{1Dprobs}
b_m = \frac{r}{r m + N-m}, \quad \quad d_m = \frac{1}{r m + N-m},
\end{equation}
where the probability $d_m$ of decreasing the mutant population by one is found by similar reasoning. Note that this derivation fails for $m=1$ ($m=N-1$) when the arc of mutants (non-mutants) contains only one node, but one can check Eq.~(\ref{1Dprobs}) still holds for these cases. These quantities play the role of transition probabilities in a Markov transition matrix. The next step is to find the eigenvalues of that matrix.

\subsection{Neutral fitness}\label{latticeNeutral}
First we work out the eigenvalues for the case of neutral fitness, $r=1$. In this case, the transition probabilities are equal, $b_m = d_m = 1/N$, and independent of $m$. Therefore, the Moran process is simply a random walk, with events occurring at a rate of $2/N$ per time step. The associated transition matrix is tridiagonal Toeplitz, which has eigenvalues given by 
\begin{equation}\label{1DneutralEigs}
\lambda_m = \frac{2}{N} - \frac{2}{N}\cos \bigg (\frac{m \pi}{N}\bigg), \quad m = 1, 2, \dots, N-1.
\end{equation}
Applying Eq.~(\ref{fixation_cumulants}) and computing the leading asymptotic form of the sums $S_n = \sum_{m=1}^{N-1} (\lambda_m)^{-n}$ (see Supplemental Material, Section S3 \cite{SM}), we find that as $N\rightarrow \infty$, the fixation-time distribution has cumulants 
\begin{equation}\label{1DneutralCumulants}
\kappa_n = (n-1)! \, \frac{\zeta(2 n)}{\zeta(4)^{n/2}},
\end{equation}
where $\zeta$ denotes the Riemann zeta function. In particular, the skew $\kappa_3 = 4 \sqrt{10}/7 \approx 1.807$, as previously calculated by Ottino-L\"{o}ffler et al.~\cite{ottino2017evolutionary} via martingale methods. The other cumulants (and characteristic function below) haven't previously been computed for the Bd process on the 1D lattice. The largeness of the skew stems from the recurrent property of the random walk. As $N \rightarrow \infty$, long walks with large fixation times become common and the system revisits each state infinitely often \cite{durrett2010probability}.

Knowledge of the cumulants allows us to obtain the exact characteristic function of the fixation-time distribution: 
\begin{equation}\label{1DneutralChar}
\phi(\omega) = e^{-\sqrt{\frac{5}{2}}\omega } \, \Gamma \left (1 - \frac{90^{1/4} \sqrt{\omega}}{\pi} \right) \Gamma \left(1+ \frac{90^{1/4} \sqrt{\omega}}{\pi} \right). 
\end{equation}
Although we cannot find a simple expression for the  distribution itself, we can efficiently evaluate it by taking the inverse Fourier transform of the characteristic function numerically. Figure~\ref{fig:ringDistributions}(a) shows that the predicted fixation-time distribution agrees well with simulations.

\subsection{Non-neutral fitness}\label{ringNonNeutral}
Next, consider $r \neq 1$ with the transition probabilities given by Eq.~(\ref{1Dprobs}). Then the eigenvalues of the transition matrix are no longer expressible in closed form. If $r$ is not too large, however, the probabilities $b_m$ and $d_m$ do not vary dramatically with $m$, the number of mutants. In particular, $b_m \sim 1/N$ for all $m$ when $N$ is large. Therefore, as a first approximation we treat the Bd process on a 1D lattice as a biased random walk with $b_m = r/(1+r)$ and $d_m = 1/(1+r)$. The eigenvalues of the corresponding transition matrix are
\begin{equation}\label{1DnonNeutralEigs}
\lambda_m = 1 - \frac{2 \sqrt{r}}{1+r}\cos \bigg (\frac{m \pi}{N}\bigg), \quad m = 1, 2, \dots, N-1.
\end{equation}
The cumulants again involve sums $S_n = \sum_{m=1}^{N-1} (\lambda_m)^{-n}$, which can be approximated in the limit $N \rightarrow \infty$ by,
\begin{equation}\label{integralApprox}
S_n \approx \frac{N}{\pi} \int_0^\pi \frac{1}{(1-2 \sqrt{r}/(1+r) \cos x)^n} \, \text{d}x.
\end{equation}
Since the integral is independent of $N$ and converges for $r\neq1$, each of the sums scales linearly: $S_n \sim N$. Thus, using Eq.~(\ref{fixation_cumulants}), we see that all cumulants past second order approach 0,
\begin{equation}
\kappa_n \sim \frac{1}{N^{(n-2)/2}} \xrightarrow{N\rightarrow\infty} 0, \quad n\geq3.
\end{equation}
Hence the fixation-time distribution is asymptotically normal, independent of fitness level. 

By evaluating the integrals in Eq.~(\ref{integralApprox}), we can more precisely compute the scaling of the cumulants. For the skew we find 
\begin{equation}\label{skewScaling}
\kappa_3 \approx \frac{2+2r(r+4)}{(r+1)\sqrt{(r^2-1)}}\frac{1}{\sqrt{N}}.
\end{equation}
The integral approximation becomes accurate when the first term in the sums $S_n$ becomes close to the value of the integrand evaluated at the lower bound ($x=0$). The fractional difference between these quantities is
\begin{equation}
\begin{split}
\Delta &= \bigg|\frac{(1-2 \sqrt{r}/(1+r))^{n}}{(1-2 \sqrt{r}/(1+r)\cos(\pi/N))^{n}} - 1 \bigg| \\
&= \frac{\sqrt{r} n \pi^2 }{(\sqrt{r}-1)^2 N^2} + \mathcal{O}(1/N^4).
\end{split}
\end{equation}
Then we have $\Delta \ll 1$ when $N \gg N_c$ where $N_c \approx 2 \pi \sqrt{n}/(r-1)$ (assuming $r$ is near 1). For the skew, we require the sums with $n = 2$ and $3$, giving $N_c \approx 10/(r-1)$.

The above calculation fails for $r \gg 1$, because when $r = \infty$ the transition probabilities $b_m = 1/m$ have different asymptotic behavior as $N\rightarrow \infty$. In particular, more time is spent waiting at states with large $m$. The process still has normally distributed fixation times \cite{ottino2017evolutionary}, but the skew becomes
\begin{equation}
\kappa_3^{\infty} = 2 \left(\sum_{m=1}^{N-1} m^3 \right ) \bigg/\left(\sum_{m=1}^{N-1} m^2 \right )^{3/2} \approx \frac{3 \sqrt{3}}{2}  \frac{1}{\sqrt{N}},
\end{equation}
for large $N$. Notice that the coefficient is different from that given by the infinite-$r$ limit of Eq.~(\ref{skewScaling}), $\kappa_3 \approx 2/\sqrt{N}$. We conjecture that there is a smooth crossover between these two scaling laws with the true skew given approximately by 
\begin{equation}\label{fullScaling}
\tilde{\kappa}_3 =  \kappa_3  \left[r^{-q} + \frac{3 \sqrt{3}}{4} (1- r^{-q}) \right]
\end{equation}
for some exponent $q$, where $\kappa_3$ is the skew given in Eq.~(\ref{skewScaling}). For small $r$ this ansatz has skew similar to that of a random walk, but captures the correct large-$r$ limit. We do not have precise theoretical motivation for this ansatz, but as discussed below, it works quite well.

\begin{figure}[t]
\includegraphics[width=\linewidth]{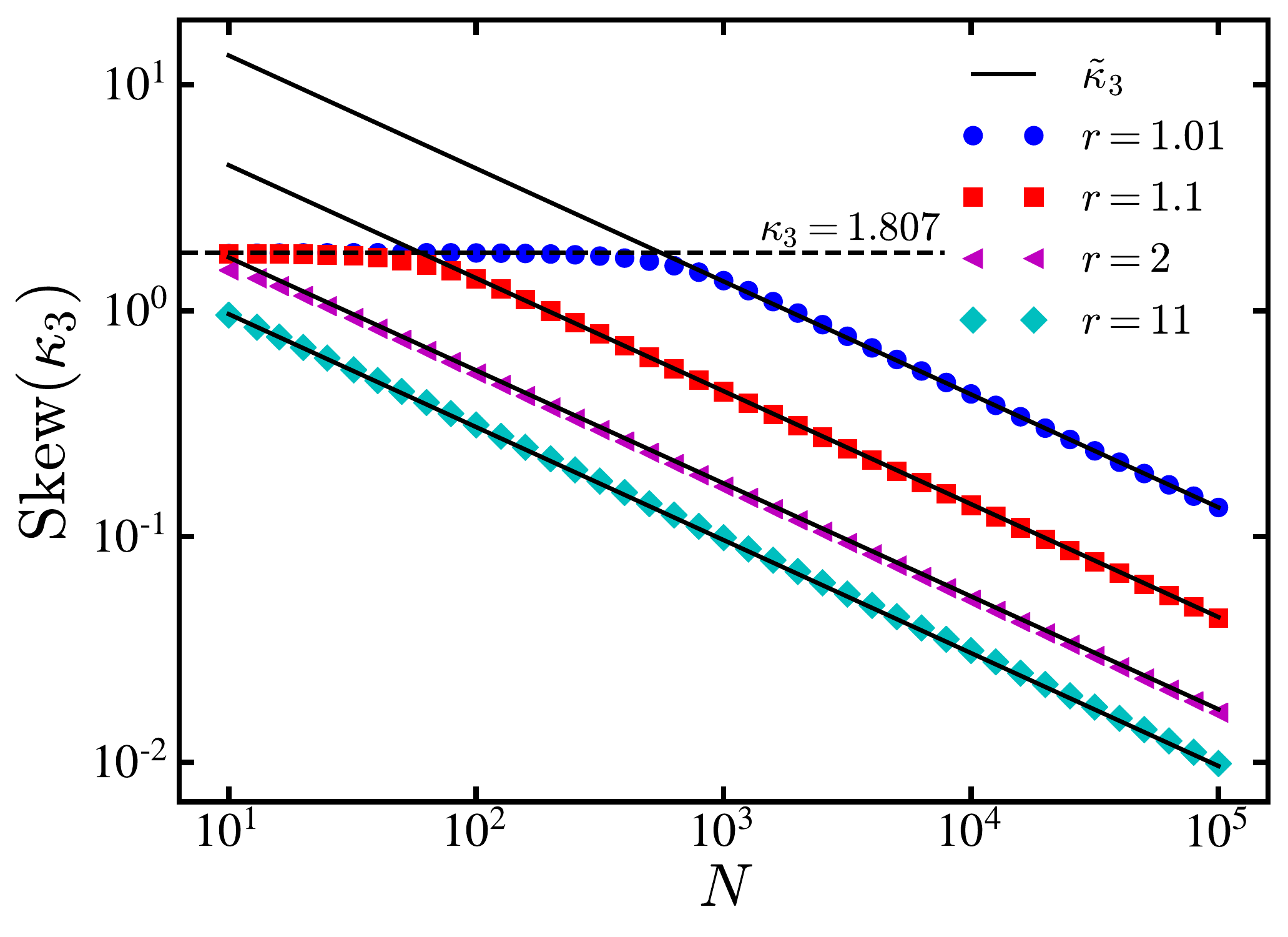}
\caption{\label{fig:ringScaling} Scaling of the skew of the fixation-time distribution on the 1D lattice with non-neutral fitness. Data points show numerical calculation of the skew for various fitness levels. The solid lines are the predicted scaling given in Eq.~(\ref{fullScaling}) with exponent $q = 1/2$ for each value of fitness $r$. For small $N$ (and small enough $r$), the skew is that of a random walk, namely $\kappa_3 = 1.807$, as shown by the dashed line. For large $N$, the skew $\kappa_3 \sim 1/\sqrt{N}$ with an $r$-dependent coefficient.}
\end{figure}

Numerical calculation of the skew for the 1D lattice was performed using the recurrence relation method discussed in Section~\ref{sec:reccurence}. The results are shown in Fig.~{\ref{fig:ringScaling}} for a few values of $r$. This calculation confirms our initial hypothesis, near neutral fitness the waiting times are uniform enough that the process is well approximated by a biased random walk and the skew approaches 0, scaling in excellent agreement with Eq.~(\ref{skewScaling}). When $N \ll N_c$, the bias is not sufficient to give the mutants a substantial advantage: the process is dominated by drift and the fixation-time distribution has large skew $\kappa_3 \approx 1.807$, as found in the preceding section. For $N \gg N_c$, selection takes over, the cumulants approach 0, and the distribution becomes normal. A similar crossover appears in the study of the fixation probability, where a transition from $\varphi_1 \sim 1/N$ to $\varphi_1 \sim 1-1/r$ is seen when $N$ passes a critical system size (that is slightly different than $N_c$). For large fitness $r \gg 1$, the ansatz Eq.~(\ref{fullScaling}) captures the scaling behavior if we use an exponent $q = 1/2$. Direct numerical simulations of the process confirm that, for any $r>1$, the fixation time on the 1D lattice has an asymptotically normal distribution [Fig.~{\ref{fig:ringDistributions}}(b)].

The random walk approximation allows us to find the asymptotic scaling of the fixation-time cumulants, but ignores the heterogeneity of waiting times present in the Moran process. Using visit statistics we can compute the cumulants exactly with Eq.~(\ref{exactCumulants}) and rigorously prove they vanish as $N\rightarrow \infty$, verifying that the waiting times have no influence on the asymptotic form of the distribution. For details, see Supplemental Material, Section S3 \cite{SM}.

Our analysis of the 1D lattice reveals an intriguing universality property of its fixation-time distribution. For any value of relative fitness $r$ other than $r=1$, the fixation-time distribution approaches a normal distribution as $N \rightarrow \infty$. Thus, for $r \neq 1$ the asymptotic shape of the distribution is universal and independent of $r$ (though bear in mind, its  mean and variance do depend on $r$). 

When $r=1$, corresponding to precisely neutral fitness, the unbiased random walk yields a qualitatively different distribution with considerably larger skew. This qualitative change as $r$ passes through unity leads to a discontinuous jump in the skew at $r=1$. 

As one might expect, the discontinuity stems from passage to the infinite-$N$ limit. For finite but large $N$, the distribution varies continuously with $r$, though our numerical results indicate that the sharp increase in skew still occurs very close to $r=1$. We will see in the next section that the discontinuity and highly skewed distribution at neutral fitness persist when we alter the network structure from a locally connected 1D lattice to a fully connected complete graph.

\section{Complete Graph}\label{completeGraph}
Next we consider the Moran process on a complete graph, useful for modeling well-mixed populations in which all individuals interact. By similar reasoning to above, given $m$ mutants the probability of adding a mutant in the next time step is
\begin{equation}\label{completeBirth}
b_m = \frac{r m}{r m + N-m} \cdot \frac{N-m}{N-1},
\end{equation}
while the probability of subtracting a mutant is
\begin{equation}\label{completeDeath}
d_m = \frac{N-m}{r m + N-m} \cdot \frac{m}{N-1}.
\end{equation}
Interestingly, as we will see in this section, these transition probabilities give rise to a fitness dependent fixation-time distribution, in stark contrast to the universality of the normal distribution observed on the 1D lattice.

\subsection{Neutral Fitness}

\begin{figure}[t]
\includegraphics[width=\linewidth]{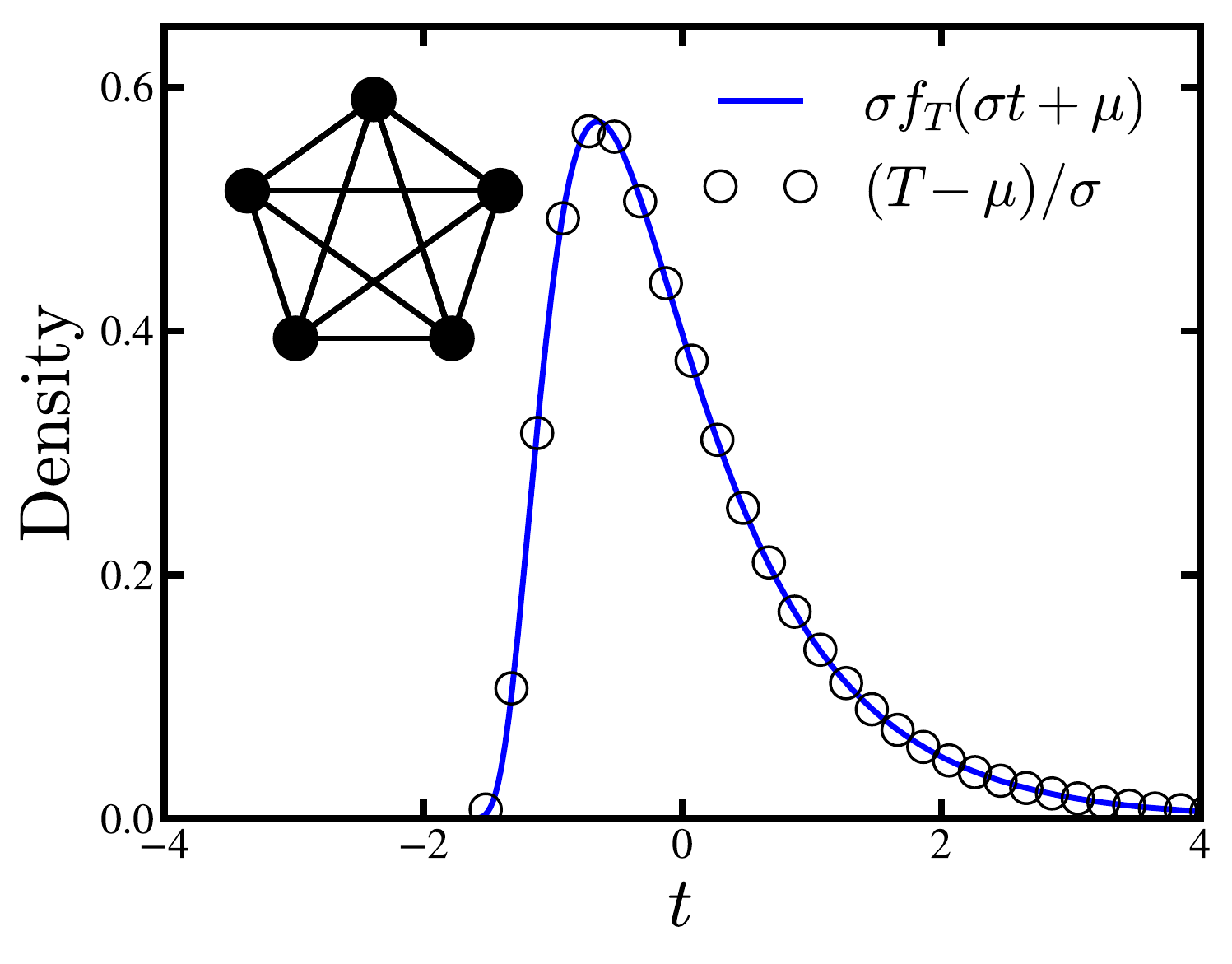}
\caption{\label{fig:completeNeutral} Fixation-time distributions on the complete graph with $N=100$ nodes and neutral fitness ($r=1$) obtained from $10^6$ simulation runs. The distribution is standardized to zero mean and unit variance. The solid curve is the theoretical distribution obtained by numerically evaluating the infinite series in Eq.~(\ref{completeDistribution}) for each value of $t$.}
\end{figure}

Again we begin with neutral fitness $r=1$. Now $b_m = d_m = (Nm-m^2)/(N^2-N)$. The eigenvalues of this transition matrix also have a nice analytical form:
\begin{equation}
\lambda_m = \frac{m (m+1)}{N(N-1)}, \quad m = 1, 2, \dots N-1.
\end{equation}
The asymptotic form of the sums $S_n = \sum_{m=1}^{N-1} (\lambda_m)^{-n}$, can be found by taking the partial fraction decomposition of $(\lambda_m)^{-n}$ and evaluating each term individually. The resulting cumulants are 
\begin{eqnarray}\label{completeCumul}
\kappa_n &=&  (n-1)!  \frac{3^{n/2}}{(\pi^2 - 9)^{n/2}} \\
& & \times \, (-1)^n \sum_{k=1}^n {2 n-k-1 \choose n-1} \left[\zeta(k) \left(1 + (-1)^{k} \right) -1\right]. \nonumber
\end{eqnarray} 
Our knowledge of the eigenvalues also allows us to obtain a series expression for the asymptotic distribution using Eq.~(\ref{fixation_distribution}). For $N\rightarrow \infty$ the standardized distribution is,
\begin{equation}\label{completeDistribution}
\begin{split}
\sigma f_T(\sigma t + \mu) = c_\sigma \sum_{j=1}^{\infty} (-&1)^{j+1} j(j+1)(2j+1) \\
& \times \exp \left[ j(j+1) (c_\sigma t+1)\right],
\end{split}
\end{equation}
where to leading order in $N$ the mean and standard deviation are $\mu = N^2$ and $\sigma = c_\sigma N^2$, with $c_\sigma = \sqrt{\pi^2/3-3}$.
This distribution was previously found using a different approach by Kimura, who also computed the first few fixation-time moments \cite{kimura1970length}. We have extended these results, obtaining the cumulants to all orders. Figure~\ref{fig:completeNeutral} shows that the predicted asymptotic distribution agrees well with numerical experiments.

\begin{figure*}[t]
\includegraphics[width=0.49\linewidth]{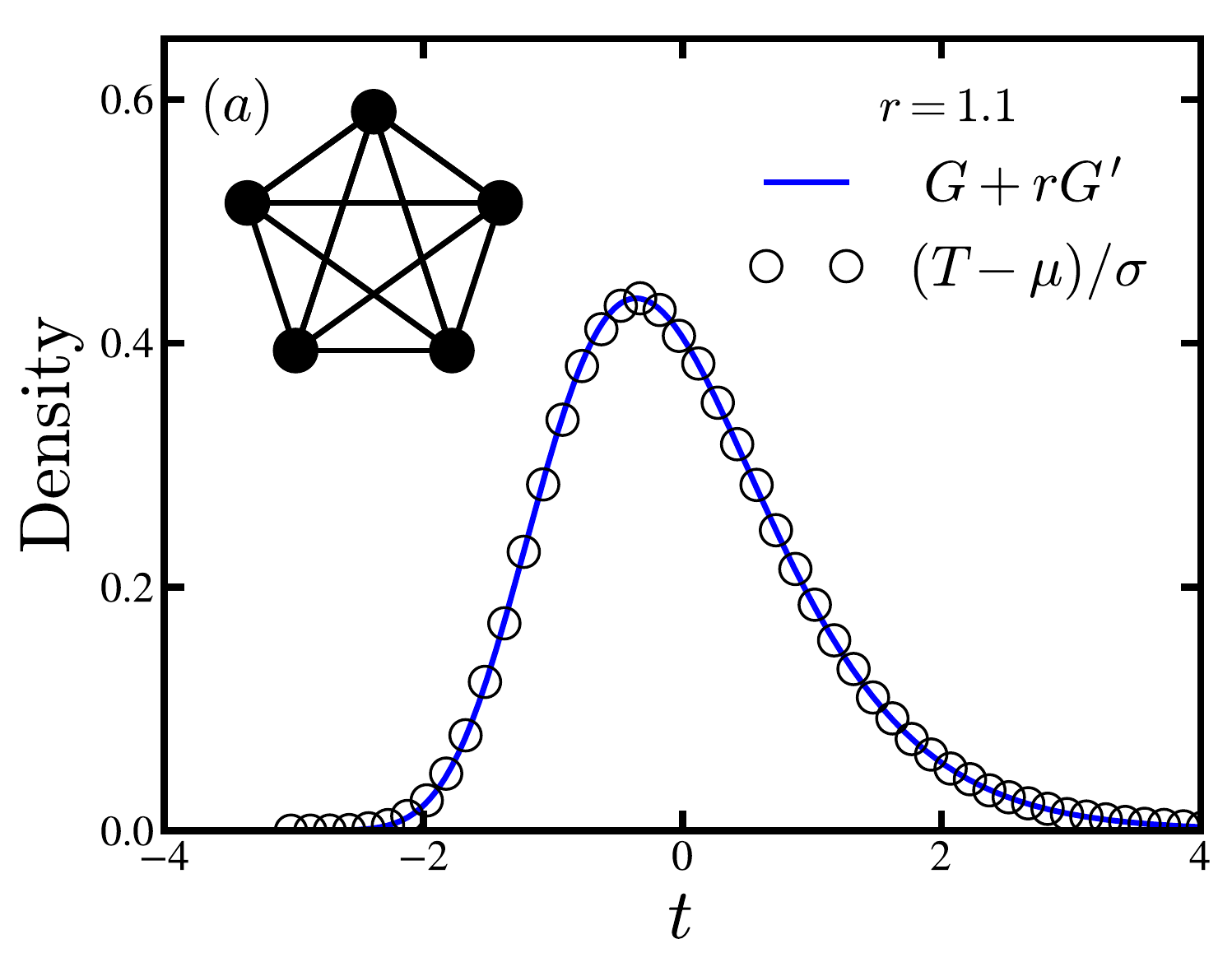}\,\,\,\, \includegraphics[width=0.49\linewidth]{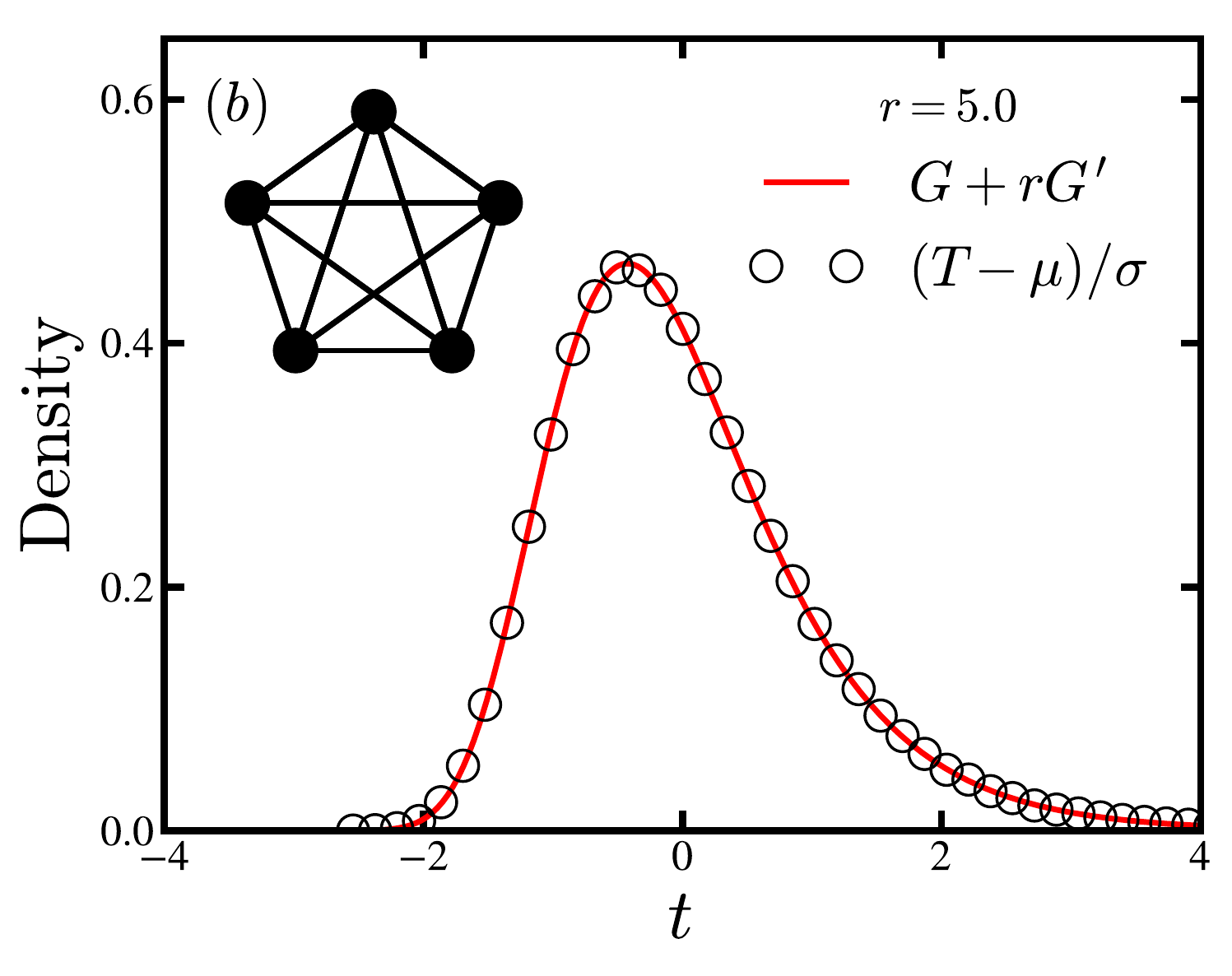}
\caption{\label{fig:completeNonNeutral} Fixation-time distributions on the complete graph with $N=5000$ nodes and non-neutral fitness ($r>1$) obtained from $10^6$ simulation runs. All distributions are standardized to zero mean and unit variance. Solid curves are the theoretical predictions obtained by numerical convolution of two Gumbel distributions, one weighted by $r$. Distributions are shown for (a) $r = 1.1$ and (b) $r = 5.0$. For larger $r$, the distribution has larger skew and a slightly sharper peak.}
\end{figure*}

The numerical value of the fixation-time skew for the Birth-death process on the complete graph is $\kappa_3 =6 \sqrt{3} (10-\pi^2)/(\pi^2-9)^{3/2}  \approx 1.6711$, slightly less than that for the 1D lattice. This decrease is the result of two competing effects contributing to the skew. First, since the birth and death transition probabilities are the same, the process is a random walk, which has a highly skewed fixation-time distribution, as shown above. The average time spent in each state, however, varies with $m$. For instance, when $m = 1 \text{ or } N-1$, $b_m \rightarrow 0$ for large $N$. But if $m = \alpha N$ for some constant $0<\alpha<1$ independent of $N$, then $b_m$ approaches a constant. 

Intuitively, the beginning and end of the mutation-spreading process are very slow because the transition probabilities are exceedingly small. To start, the single mutant must be selected by chance to give birth from the $N$ available nodes, a selection problem like finding a needle in a haystack. Similarly, near fixation the reproducing mutant must find and replace one of the few remaining non-mutants, again choosing it by chance from an enormous population.

The characteristic slowing down at certain states is reminiscent of ``coupon collection'', as discussed earlier. Erd\H{o}s and R\'{e}nyi proved that for large $N$, the normalized time to complete the coupon collection follows a Gumbel distribution \cite{erdos1961classical}, which we denote by $\text{Gumbel}(\alpha, \beta)$ with density 
\begin{equation}
f(t) = \beta^{-1} e^{-(t-\alpha)/\beta} \exp(-e^{-(t-\alpha)/\beta}).
\end{equation}
For the Moran process, each slow region is produced by long waits for the random selection of rare types of individuals: either mutants near the beginning of the process or non-mutants near the end. In the next section we show that the two coupon collection regions of the Bd process on a complete graph lead to fixation-time distributions that are convolutions of two Gumbel distributions. In the case of neutral fitness, these Gumbel distributions combine with the random walk to produce a new highly skewed distribution with cumulants given by Eq.~(\ref{completeCumul}).

\subsection{Non-neutral fitness}\label{completeNonNeutralSec}
 
We saw in Section \ref{ringNonNeutral} that when the average time spent in each state is constant or slowly varying the fixation-time distribution is asymptotically normal. Birth-death dynamics on the complete graph, however, exhibit coupon collection regions at the beginning and end of the process, where the transition probabilities vanish. We begin this section with a heuristic argument that correctly gives the asymptotic fixation-time distribution in terms of independent iterations of coupon collection.

Differentiating $b_m$ with respect to $m$, we find the slope near $m=0$ is $(r+1)/N$, while the slope near $m=N$ has magnitude $(r+1)/(r N)$ for $N \gg 1$. The transition rates approach zero at each of these points, so we expect behavior similar to coupon collection giving rise to two Gumbel distributions. Since the slope is greater for $m$  near 0 than for $m$ near $N$, the Moran process completes its coupon collection faster near the beginning of the process than near fixation. 

This heuristic suggests that the asymptotic fixation time should be equal in distribution to the sum of two Gumbel random variables, one weighted by $r$, which is the ratio of the slopes in the coupon collection regions. Specifically, if $T$ is the fixation time with mean $\mu$ and variance $\sigma^2$, we expect
\begin{equation}\label{completeFitDist}
\frac{T -\mu}{\sigma} \xrightarrow{d} \frac{G + r G'}{\sqrt{1+r^2}},
\end{equation}
where $\xrightarrow{d}$ means convergence in distribution for large $N$. Here $G$ and $G'$ denote independent and identically distributed Gumbel random variables with zero mean and unit variance. It is easy to check that the correct distribution is $\text{Gumbel}(-\gamma \sqrt{6}/\pi, \sqrt{6}/\pi)$, where $\gamma \approx 0.5772$ is the Euler-Mascheroni constant.

Let us make this argument more rigorous. Previous theoretical analysis showed that in the infinite fitness limit, the fixation time has an asymptotically Gumbel distribution \cite{ottino2017evolutionary}. This result can be recovered within our framework, since when $r = \infty$ it follows that $d_m = 0$, so the eigenvalues of the transition matrix are just $\lambda_m = b_m = (N-m)/(N-1)$ and the cumulants can be directly calculated using Eq.~(\ref{fixation_cumulants}). 

For large (but not infinite) fitness, the number of mutants is monotonically increasing, to good approximation, since the probability that the next change in state increases the mutant population is $r/(1+r) \approx 1$. The time spent waiting in each state, however, changes dramatically, especially near $m=1$. Here, $b_1\rightarrow 0$ for large $N$, in stark contrast to the infinite fitness system where $b_1 \rightarrow 1$. The time spent at each state, $t_m$ is an exponential random variable, $\mathcal{E}(b_m+d_m)$. In this approximation each state is visited exactly once, so the total fixation time is a sum of these waiting times: 
\begin{equation}
T \approx \sum_{m=1}^{N-1} \mathcal{E}(b_m + d_m). 
\end{equation}
But this sum of exponential random variables has density given by Eq.~(\ref{fixation_distribution}), with the substitution $\lambda_m \rightarrow b_m + d_m$. Thus, the cumulants of $(T-\mu)/\sigma$ are
\begin{equation}\label{completeFitCumul}
\kappa_n = \frac{1+r^n}{(1+r^2)^{n/2}} \times \frac{(n - 1)! \zeta(n)}{\zeta(2)^{n/2}},
\end{equation}
which are exactly the cumulants corresponding to the sum of Gumbel random variables given in Eq.~(\ref{completeFitDist}). In the limit $r \rightarrow \infty$, the first term in Eq.~(\ref{completeFitCumul}) becomes $1$, and the cumulants are those for a single Gumbel distribution, in agreement with previous results \cite{ottino2017evolutionary}.

\begin{figure}[t]
\includegraphics[width=\linewidth]{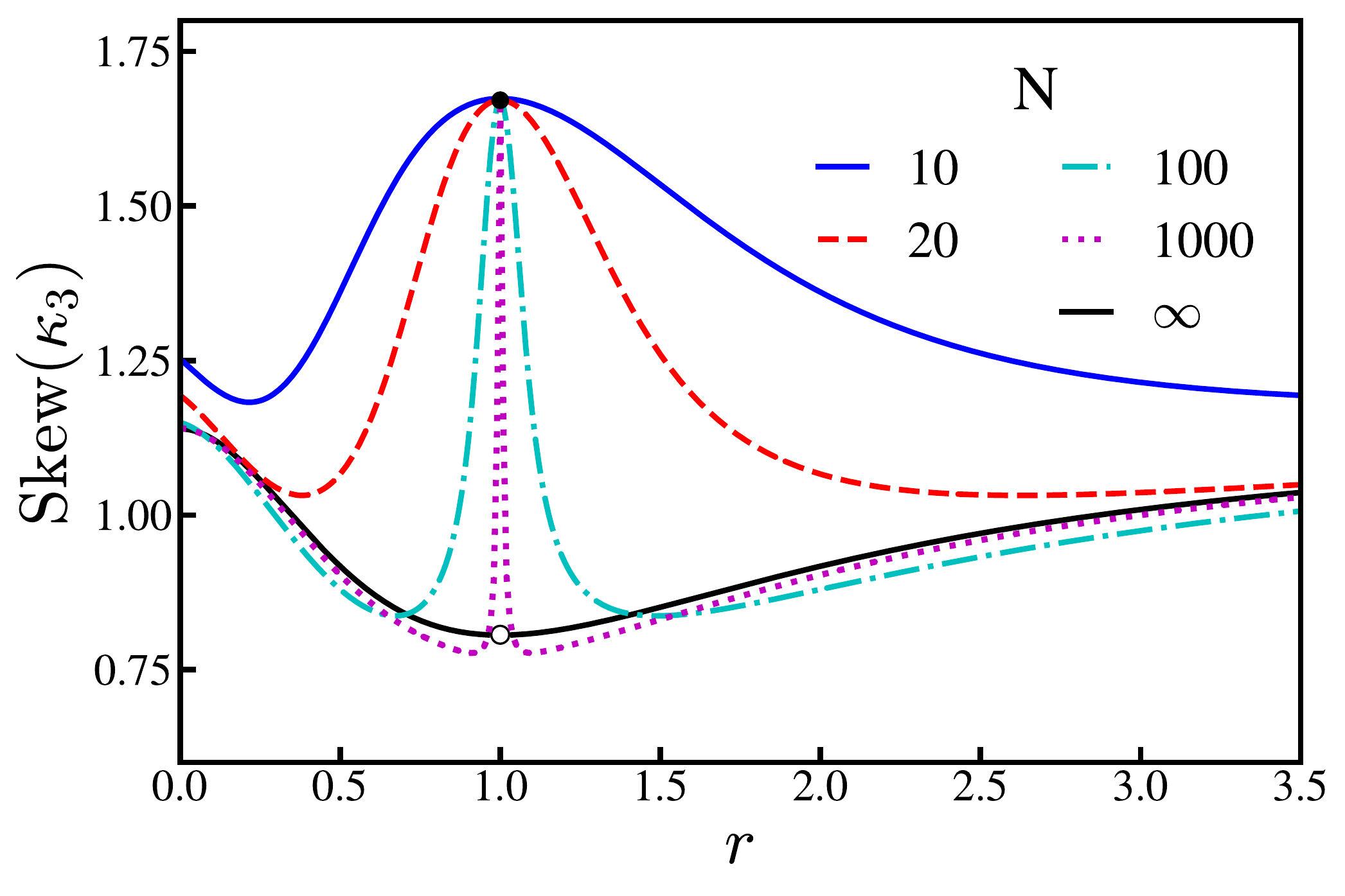}
\caption{\label{fig:completeFiniteN} Fitness dependence of fixation-time skew for the Moran Birth-death process on the complete graph. The skew is shown for $r\geq0$ and is invariant under $r \rightarrow 1/r$. For finite $N$, the skew does not have a discontinuity, but does show non-monotonic dependence on fitness $r$. In particular, for a given $N$, there is a certain fitness level with minimum skew. As $N \rightarrow \infty$, we see non-uniform convergence to the predicted skew given by $\kappa_3$ in Eq.~(\ref{completeFitCumul}), leading to the discontinuity at $r=1$. Moreover, for fixed $r$, the convergence to the $N=\infty$ skew is non-monotonic.}
\end{figure}

Remarkably, these cumulants are exact for any $r>1$, not just in the large-$r$ limit. We can see this directly for the skew $\kappa_3$ using the visit statistics approach, computing the asymptotic form of Eq.~(\ref{exactCumulants}) with the complete graph transition probabilities, Eqs.~(\ref{completeBirth}) and (\ref{completeDeath}). Details of the asymptotic analysis are provided in Supplemental Material, Section S4 \cite{SM}. Numerical simulations of the Moran process corroborate our theoretical results. As shown in Fig.~\ref{fig:completeNonNeutral}, for $r=1.1$ and $r=5$ the agreement between simulated fixation times and the predicted convolution of Gumbel distributions is excellent, at least when $N$ is sufficiently large. Again, our calculations show a discontinuity in the fixation-time distribution at $r=1$. In particular, the $r\rightarrow 1$ limit of the cumulants for non-neutral fitness in Eq.~(\ref{completeFitCumul}) is not the same as the cumulants for neutral fitness found in the preceding section [Eq.~(\ref{completeCumul})]. 

For smaller networks, it is fascinating to see how the results converge to the asymptotic predictions as $N$ grows. Figure~\ref{fig:completeFiniteN} shows how the skew of the fixation-time distribution depends on $r$ and $N$ for the complete graph. As discussed in Section~\ref{disadvantageousMutants}, the fixation-time distributions for these systems are invariant under $r \rightarrow 1/r$. Therefore we show the skew for all $r>0$, to emphasize the intriguing behavior near neutral fitness, where $r=1$. We find that non-uniform convergence of the fixation-time skew leads to the discontinuity predicted at $r=1$. For finite $N$, the skew is a non-monotonic function of $r$ and has a minimum value at some fitness $r_\text{min}(N)$. Furthermore, at fixed $r$, the convergence to the $N=\infty$ limit is itself non-monotone. Though beyond the scope of the current study, further investigation of this finite-$N$ behavior would be worth pursuing. 

\section{Partial fixation times}\label{truncationSection}

In many applications, we may be interested in the time to partial fixation of the network. For instance, considering cancer progression \cite{williams1972stochastic, sottoriva2013intratumor, bozic2013evolutionary} or the incubation of infectious diseases \cite{ottino2017evolutionary}, symptoms can appear in a patient even when a relatively small proportion of cells are malignant or infected. We therefore consider $T_\alpha$, the total time to first reach $\alpha N$ mutants on the network, where $0<\alpha<1$. The methods developed in Section~\ref{generalTheory} apply to these processes as well. For the eigendecomposition approach we instead use the sub-matrix of $\Omega_\text{tr}$ containing the first $\alpha N$ rows and columns. In calculations involving the numerical recurrence relations or visit statistics, we simply cut the sums off at $\alpha N$ instead of $N$ and for the latter, replace $w_{i_1i_2\cdots i_n}(r, N)$ with $w_{i_1i_2\cdots i_n}(r, \alpha N)$.

\subsection{One-dimensional lattice}
Truncating the Moran Bd process on the 1D lattice by a factor $\alpha$ has no effect on the asymptotic shape of the fixation-time distributions. In both the neutral fitness system and the random walk approximation to the non-neutral fitness system, the transition matrix has no explicit dependence on the state or system size [aside from proportionality factors that cancel in Eq.~(\ref{fixation_cumulants})]. Thus, the eigenvalues are identical to those calculated previously, but correspond to a smaller effective system size $\alpha N$. Taking the limit $N \rightarrow \infty$ therefore yields the same asymptotic distributions found in Section~\ref{1D_lattice}.

\begin{figure*}[t]
\includegraphics[width=0.485\linewidth]{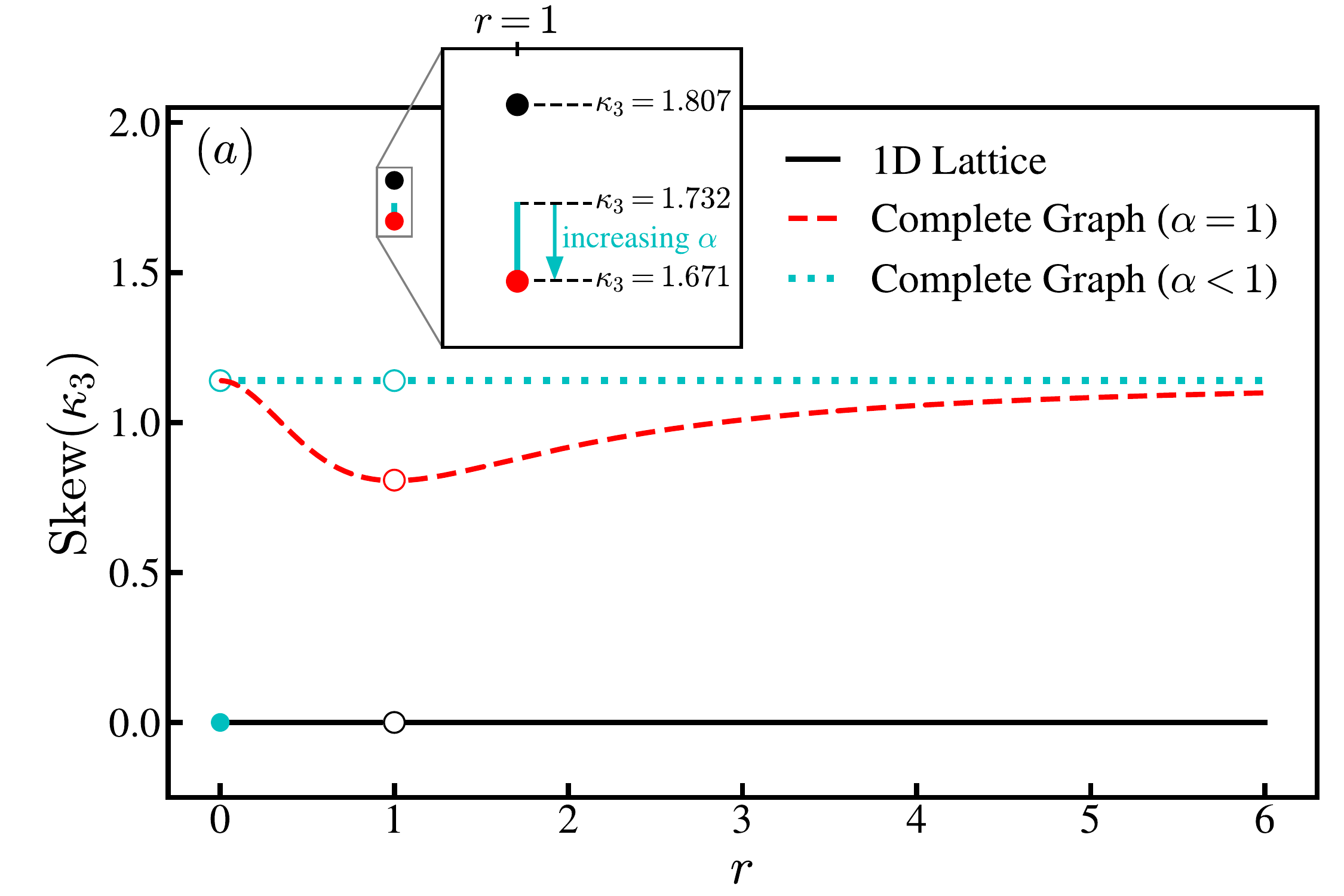} \,\,\,\, \includegraphics[width=0.485\linewidth]{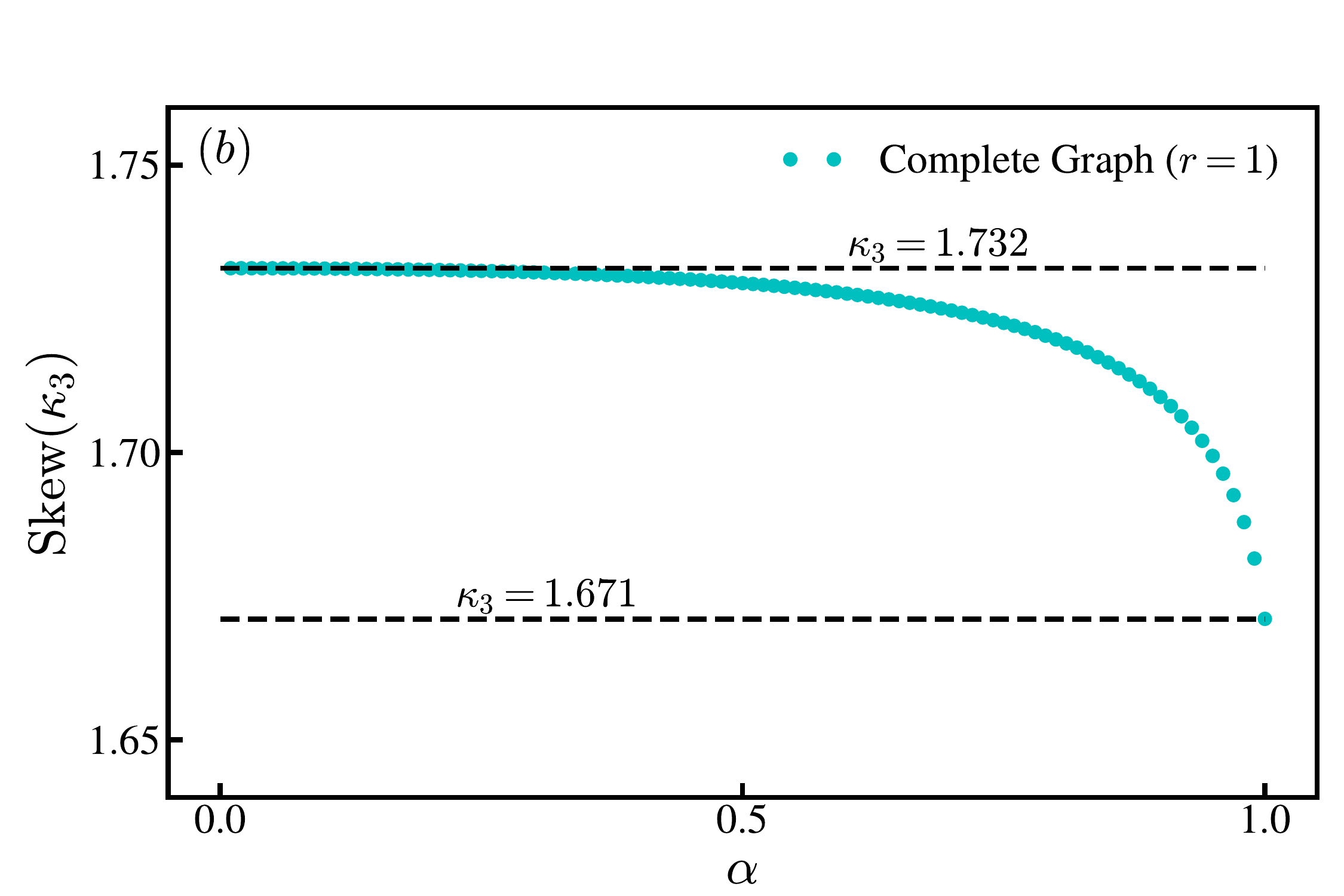}
\caption{\label{fig:fitnessDependenceSkew} Variation of fixation-time skew $\kappa_3$ with fitness level $r$ and truncation factor $\alpha$ for different network structures. (a) The skew of the fixation-time distribution is plotted versus fitness for the 1D lattice (black solid line), complete graph (red dashed line), and complete graph with truncation (green dotted line). The skew is shown for all $r\geq0$ and is invariant under $r \rightarrow 1/r$.  When $r\neq 1$ and $r<\infty$, the fixation-time distribution is normal for the 1D lattice, and hence has zero skew ($\kappa_3=0$). The distribution becomes a fitness-weighted convolution of Gumbel distributions for the complete graph, and a single Gumbel distribution for the complete graph with truncation (for any $\alpha<1$). Each curve jumps discontinuously at $r=1$, where the distributions become highly skewed with $\kappa_3 > 1.5$. The inset shows a blow-up of the neutral fitness results, specifying the skew for each case. On the complete graph with truncation, the skew  is continuously variable at $r=1$, taking on an interval of values between $6 \sqrt{3} (10-\pi^2)/(\pi^2-9)^{3/2} \approx 1.671$ when $\alpha=1$, and $\sqrt{3}\approx 1.732 $ when $\alpha = 0$. This range is indicated by the green vertical line. The truncated fixation time on the complete graph has a second discontinuity at $r=\infty$ (shown here at $r=0$, by exploiting the $r \rightarrow 1/r$ invariance). At this discontinuity the functional form of the distribution jumps from Gumbel to normal. (b) The fixation-time skew for the complete graph with neutral fitness, plotted versus the truncation factor $\alpha$. These points correspond to the green vertical line in panel (a) at $r=1$.}
\end{figure*}

\subsection{Complete graph: truncating coupon collection}

The complete graph exhibits more interesting dependence on truncation. Since the transition probabilities have state dependence, the eigenvalues change with truncation (they don't correspond to the same system with smaller effective $N$). Our intuition from coupon collection, however, lets us predict the resulting distribution.

First consider non-neutral fitness. Then there are two coupon collection stages, one near the beginning and another near the end of the process, and  together they generate a fixation-time distribution that is a weighted convolution of two Gumbel distributions. The effect of truncating the process near its end should now become clear: it simply removes the second coupon collection. The truncated process stops before the mutants have to laboriously find and replace the last remaining non-mutants. Therefore, we intuitively expect the fixation time for non-neutral fitness to be distributed according to a single Gumbel distribution, regardless of fitness level.

The only exception occurs if $r=\infty$; then no coupon collection occurs at the beginning of the process either, as the lone mutant is guaranteed to be selected to give birth in the first time step, thanks to its infinite fitness advantage. Thus, when fitness is infinite and the process is truncated at the end, both coupon collection phases are removed and the fixation times are normally distributed.

Similar reasoning applies to the Birth-death process with neutral fitness. It also has two coupon collection regions, one of which is removed by truncation. In this case, however, the random walk mechanism contributes to the skew of the overall fixation-time distribution, combining non-trivially with the coupon collection-like process. We find that the skew of the fixation time  depends on the truncation factor $\alpha$, varying between $6 \sqrt{3} (10-\pi^2)/(\pi^2-9)^{3/2} \approx 1.6711$ when $\alpha=1$, and $\sqrt{3}\approx 1.732 $ when $\alpha = 0$. A derivation of this $\alpha\rightarrow 0$ limit is given in Supplemental Material, Section S4 \cite{SM}.

\subsection{Summary of main results}

The main results from Sections~\ref{1D_lattice}--\ref{truncationSection} are summarized in Fig.~\ref{fig:fitnessDependenceSkew}, which shows the asymptotic fitness dependence of fixation-time skew for each network considered in this paper. We again show the skew for all $r>0$ (not just $r > 1$) to emphasize the discontinuities at zero, neutral, and infinite fitness. On the 1D lattice, independent of the truncation factor $\alpha$, the Bd process has normally distributed fixation times, except at neutral fitness where the distribution is highly skewed. The complete graph fixation-time distributions are the weighted convolution of two Gumbel distributions for $r\neq1$, again with a highly skewed distribution at $r=1$. With truncation by a factor $\alpha<1$, the distribution for the complete graph is Gumbel for $1<r<\infty$, and normal for $r=\infty$. With neutral fitness the fixation distribution is again highly skewed, with skew dependent on the truncation factor $\alpha$.

\section{Extensions}\label{discussion}

It is natural to ask whether our results are generic; do the same fixation-time distributions appear in other models of evolutionary dynamics? Here we explore the robustness of our results to various changes in the model update dynamics and the network topology. The main finding is that our results are insensitive to these changes, at least qualitatively. The distributions typically remain right-skewed and even follow the same functional forms derived above.

\subsection{Other update dynamics}

\subsubsection{ Two-fitness Moran process}
The Moran Bd processes considered above require a single fitness level, designating the relative reproduction rates between mutants and non-mutants. Another common model is the Moran Birth-Death (BD) process \cite{Note1}, which has a second fitness level $\tilde{r}$ measuring the resilience of mutants versus non-mutants during the replacement step \cite{kaveh2015duality}. Taking this into account, when a mutant or non-mutant is trying to replace its neighbors, mutants are replaced with probability proportional to $1/\tilde{r}$. Taking $\tilde{r} = 1$ returns to the model used throughout the preceding sections. The two-fitness model may better capture the complexity of real-world evolutionary systems but does not generally give rise to qualitatively different fixation-time distributions. For brevity, we simply discuss the resulting fixation-time distributions for the BD model. Details supporting the results quoted below are provided in the Supplemental Material, Section S5 \cite{SM}.

Writing down the transition probabilities for the Moran BD process, we find that $b_m/d_m \rightarrow r \tilde r$ as $N \rightarrow \infty$. This motivates the definition of an effective fitness level, $r_\text{eff}=r \tilde r$. When $r_\text{eff}\neq 1$ our results from above translate to this model. On the 1D lattice the fixation times are normally distributed, while on the complete graph the fixation time distribution is a weighted convolution of Gumbel distributions $G + (r/\tilde r) G'$, with relative weighting $r/\tilde r$ (instead of $r$). When $r_\text{eff}=1$, the process is asymptotically unbiased and we expect a highly skewed fixation-time distribution. This is indeed the case, although numerical calculations indicate there is an entire family of distributions, dependent on $r=1/\tilde r$.

It is interesting to contrast the above observations with a result in evolutionary dynamics known as the isothermal theorem. The theorem states that for $\tilde{r}=1$, the Moran process on a large class of networks, known as isothermal graphs, has fixation probability identical to the complete graph \cite{lieberman2005evolutionary}. Recent work has shown that this breaks down if  $\tilde{r}\neq1$; the fixation probability develops new network dependence \cite{kaveh2015duality}. In contrast, even isothermal graphs (including the complete graph and 1D lattice) have fixation-time distributions that depend on network structure. The two-fitness BD model breaks the universality in fixation probabilities predicted by the isothermal theorem, but leads to the same family of fixation distributions that arise due to network structure.

\subsubsection{The Death-Birth Moran process}
A two-fitness Death-Birth (DB) Moran process \cite{Note1} is also frequently used to study evolutionary dynamics. In this model, the birth and death events are reversed in order. At each time step a node is chosen at random, with probability proportional to $1/\tilde r$, and one of its neighbors is chosen with probability proportional to $r$. The first individual dies and is replaced by an offspring of the same type as the neighbor. The process continues until the mutation either reaches fixation or goes extinct.

The BD and DB processes obey a duality property \cite{kaveh2015duality}. Starting from the BD transition probabilities, if we swap the two fitness levels $r \leftrightarrow \tilde r$ and substitute $m \rightarrow N-m$ (which swaps mutants and non mutants), we obtain the DB transition probabilities. Therefore, the transition matrix for the DB model is identical to that for the corresponding dual BD process, but has the main-, super-, and sub-diagonal entries reversed in order. This leaves the matrix eigenvalues unchanged, so that the DB process has identical fixation-time distributions to those given in the preceding section for the dual BD process. 

In principle, the correspondence between DB and BD fixation times could break down for the truncated process considered in Section~\ref{truncationSection}. In practice, however, the results are again generally identical. For the truncated DB process, the fixation times on the 1D lattice remain normally distributed. On the complete graph, one of two coupon collection regions is removed by truncation leading to fixation-times following a single Gumbel distribution. 

One exception, where the dual models yield different results under truncation, is at infinite fitness. As in Section~\ref{truncationSection}, at infinite fitness ($r \rightarrow \infty$) the BD model performs a single coupon collection near fixation, which is cut off by truncation, leading to a normal fixation-time distribution. In contrast, in the dual infinite-fitness DB model ($\tilde r \rightarrow \infty$) the coupon collection occurs at the beginning of the process and even under truncation the Gumbel fixation-time distribution is preserved. This effect was previously observed by Ottino-L\"{o}ffler et al. \cite{ottino2017evolutionary}.

\subsection{Other networks: Approximate results via mean-field transition probabilities}
While the 1D lattice and complete graph provide illustrative exactly solvable models of the fitness dependence of fixation-time distributions, other networks may be more realistic. On more complicated networks the analytical tools developed here fail because the transition probabilities (the probability of adding or subtracting a mutant given the current state) depend on the full configuration of mutants, not just the number of mutants. Such systems can still be modeled as a Markov process, but the state space becomes prohibitively large. Fortunately, for certain networks the effect of different  configurations can be averaged over, giving a mean-field approximation to the transition probabilities. This approach has been used on a variety of networks to calculate fixation times \cite{hajihashemi2019fixation, ying2018mean, ottino2017evolutionary, ottino2017takeover}. In this section we discuss how such mean-field approaches can be used to calculate  fixation-time distributions for evolution on several different networks.

\begin{figure}[t] 
\includegraphics[width=\linewidth]{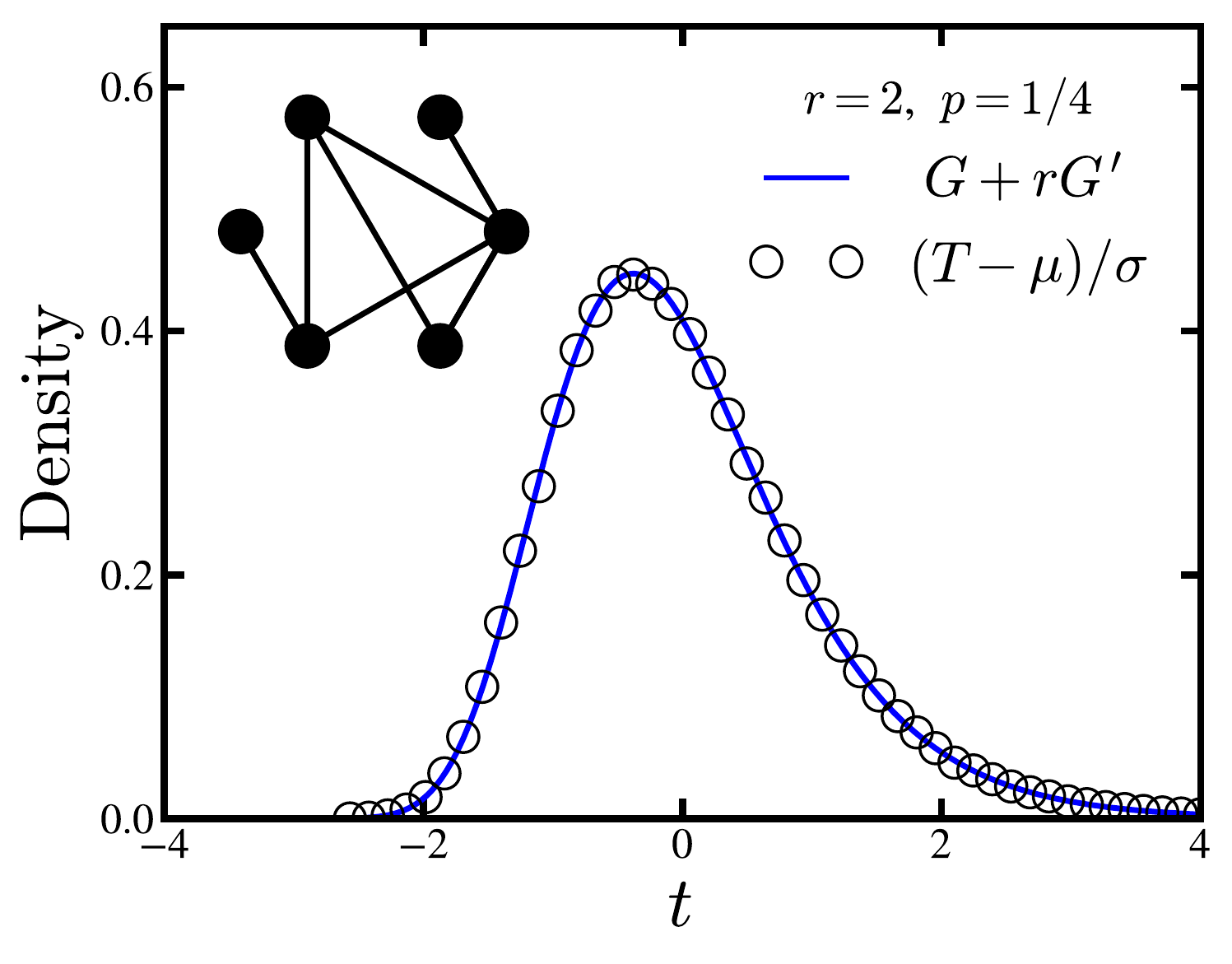}
\caption{\label{fig:ERsimulation} Fixation-time distribution on an Erd\H{o}s-R\'{e}nyi random graph with $N=100$ nodes, edge probability $p=1/4$, and fitness $r=2$, obtained from $10^6$ simulation runs (the same graph is used for each run). The distribution is standardized to zero mean and unit variance. The solid curve is the theoretical prediction for the complete graph, obtained by numerical convolution of two Gumbel distributions, one weighted by $r$. For these parameters, the random graph fixation time is captured by the mean field approximation.}
\end{figure}

\subsubsection{Erd\H{o}s-R\'{e}nyi random graph}
We start with the  Erd\H{o}s-R\'{e}nyi random graph, for which the mean-field transition probabilities were recently estimated \cite{hajihashemi2019fixation}. The result is identical to the complete graph probabilities [Eqs.~(\ref{completeBirth})-(\ref{completeDeath})] up to a constant factor $1-2/Np$, which depends on the edge probability $p$ for the network. This correction is important for computing the mean fixation time, but does not affect the shape of the fixation-time distribution, since proportionality factors cancel in Eq.~(\ref{fixation_cumulants}). Therefore we expect the asymptotic fixation-time distribution will be a weighted sum of two Gumbel distributions. This prediction holds for infinite fitness, where the fixation time on an Erd\H{o}s-R\'{e}nyi network has a Gumbel distribution \cite{ottino2017evolutionary}.

Preliminary simulations show that the Erd\H{o}s-R\'{e}nyi network has the expected fixation-time distributions for $p=1/4$ and $r=2$ (see Figure~\ref{fig:ERsimulation}). Further investigation is required to determine the range of fitness and edge probabilities for which this result holds asymptotically (as $N\rightarrow \infty$). For constant $p$, the average degree is proportional to the system size $\langle k \rangle = p N$, similar to the complete graph. It may be, however, that for some $p$ and $r$ the mean-field approximation is not sufficient to capture the higher-order moments determining the shape of the distribution. It is also traditional to consider $N$-dependent edge probabilities with $p(N)$ chosen, for example, to fix $\langle k \rangle$. It is unclear whether such graphs will behave like the ring (due to their sparsity), like the complete graph (due to their short average path length), or somewhere in between these extreme cases. In the same vein, which other networks admit accurate mean-field approximations to the transition probabilities? Do many complex networks have fixation-time distributions identical to the complete graph?

\subsubsection{Stars and superstars: evolutionary amplifiers}
Another nice approximation maps the Moran process on a star graph, a simple amplifier of selection, onto a birth-death Markov chain \cite{frean2013effect}. The resulting transition probabilities exhibit coupon collection regions, similar to the complete graph. The ratio of slopes near these regions (few mutants or non-mutants), however, is $r^2$. Our heuristic predicts the fixation-time distribution on the star is $G+r^2G'$. In addition to amplifying fixation probability, the star increases fixation-time skew. This raises a broader question: do evolutionary amplifiers also amplify fixation-time skew? Computing fixation times for evolutionary dynamics on superstars (which more strongly amplify selection \cite{lieberman2005evolutionary}) remains an open problem.

\subsubsection{Growth of cancerous tumors: evolutionary dynamics on $d$-dimensional lattices}
Mean-field arguments have also been applied to $d$-dimensional lattices in the infinite-fitness limit \cite{ottino2017evolutionary, ottino2017takeover}. In this limit the mutant population grows in an approximately spherical shape near the beginning of the process and the population of non-mutants is approximately spherical near fixation. The surface area to volume ratio of the $d$-dimensional sphere gives the probability of adding a mutant. With finite fitness, non-mutants can now replace their counterparts and the surface of the sphere of growing mutants roughens \cite{williams1972stochastic}. For near-neutral fitness, the configuration of mutants resembles the shape of real cancerous tumors. Perhaps mean-field approaches can draw connections between the fitness-dependent roughness of growing mutant populations and fixation-time distributions for evolution on lattices.

\begin{table*}
\caption{\label{summaryTable} Asymptotic fixation-time statistics for the Moran Birth-death and Death-birth processes on the complete graph and the 1D lattice. Together with the mean and variance, the standardized distributions give a complete statistical description of the fixation time. The mean and variance given are to leading order in $N$ for each case.}
\begin{ruledtabular}
\begin{tabular}{ccccc}
 &&\multicolumn{3}{c}{Asymptotic Fixation-Time Statistics}\\
  Network& Fitness Level & Mean & Variance & Standardized Distribution \\  \hline \\[-1.5ex]
  \multirow{2}{*}[-2ex]{1D Lattice}& $r=1$& $\displaystyle \frac{1}{6}N^3$ & $\displaystyle \frac{1}{90} N^6$ & Highly Skewed [Eqs.~(\ref{1DneutralCumulants}) \& (\ref{1DneutralChar})]\\  \\[-1.5ex]
& $r>1$ & $\displaystyle \frac{r+1}{2(r-1)} N^2$& $\displaystyle \frac{(r+1)(r^2+r+1)}{3(r-1)^3} N^3$ &$\mathcal{N}(0,1)$\\  \\[-1.5ex]
\multirow{2}{*}[-2ex]{Complete Graph}& $r=1$&$N^2$& $\displaystyle \left( \frac{\pi^2}{3}-3 \right )N^4$ & Highly Skewed [Eqs.~(\ref{completeCumul}) \& (\ref{completeDistribution})]\   \\  \\[-1.5ex]
&$r>1$& $\displaystyle \frac{r+1}{r-1} N \log N$ &$\displaystyle \frac{\pi^2 (r+1)^2}{6(r-1)^2}N^2$ & $G + r G'$\\
\end{tabular}
\end{ruledtabular}
\end{table*}

\section{Summary}

In this paper we have obtained the first closed-form solutions for the fitness dependence of fixation-time distributions of the Moran Birth-death process on the 1D lattice and complete graph. Previous analyses were restricted to the limit of infinite fitness, with some partial results for neutral fitness. To reiterate our new results: There is a dichotomy between neutral and non-neutral fitness. When fitness is neutral, the distribution always exhibits a discontinuity; whether the graph is complete or a 1D lattice, the skew jumps up discontinuously in either case. On the other hand, when fitness is non-neutral but otherwise arbitrary, the results depend strongly on network topology. Specifically, on the complete graph the fixation-time distribution is a fitness-weighted convolution of Gumbel distributions and hence is always skewed, whereas on the 1D lattice the distribution is normal and hence is never skewed.

Together with the mean and variance, the distributions derived here give a complete statistical description of the asymptotic fixation time (see Table~\ref{summaryTable}). Our analysis revealed that these results are robust in the sense that similar distributions arise under truncation, in some other models, and in some other network structures, including the Erd\H{o}s-R\'{e}nyi random graph.

\section{Future Directions}

Though the model we have focused on here (the Moran Birth-death model) is deliberately simplified, we expect our results will be useful in applications. For instance, the theory should allow a more refined analysis of the rate of evolution, by extending the seminal work by Kimura, whose neutral theory of evolution predicted a molecular clock \cite{kimura1968evolutionary}. In his model, neutral mutations become fixed at a constant rate, independent of population size. This result, with some refinements, is now used widely in estimating evolutionary time scales \cite{kumar2005molecular}. The fixation-time distributions discussed here should allow one to go beyond Kimura's classic analysis to capture the full range of evolutionary outcomes, by providing information about the expected deviations from the constant-rate molecular clock, as well as how this prediction is affected by population structure. More generally, it would be interesting to study the implications of these distributions for rates of evolution at various fitness levels. 

Furthermore, our results provide concrete predictions that are testable via bacterial evolution experiments. Does the same fitness and network structure dependence of fixation-time distributions arise in real systems?

Future theoretical studies could analyze random networks and lattices more deeply, as well as stars and superstars, the prototypical evolutionary amplifiers \cite{lieberman2005evolutionary}. More sophisticated models involving evolutionary games are also of interest. These have skewed fixation-time distributions \cite{ashcroft2015mean} whose asymptotic form remains unknown. Finally, we hope that methods developed here will prove useful in other areas, such as epidemiology \cite{doering2007asymptotics}, ecology \cite{doering2005extinction}, and protein folding \cite{zwanzig1992levinthal}, where stochastic dynamics may similarly give rise to skewed first-passage times.

\section*{Acknowledgments}
We thank Bertrand Ottino-L\"{o}ffler and Jacob Scott for helpful discussions. Research supported by the Cornell Presidential Life Science Fellowship and NSF Graduate Research fellowship grant DGE-1650441 to D. H. and by NSF grants DMS-1513179 and CCF-1522054 to S.H.S.

\appendix*

\section{Visit statistics}\label{appendix}

In this Appendix we formulate the visit statistics approach. We first provide further details in the derivation of the series expression for the fixation-time cumulants given in Eq.~(\ref{exactCumulants}), and then explicitly compute the weighting factors that appear in this expression to third order. This result requires constant selection, $b_m/d_m = r$, as is the case for the Moran process. Under constant selection the transient transition matrix can be written as $\Omega_\text{tr} = \Omega_\text{RW} D$, where $D$ is diagonal with elements $D_{mm} = b_m + d_m$ and $\Omega_\text{RW}$ is the transition matrix for a random walk,
\begin{equation}
    [\Omega_\text{RW}]_{nm} = \frac{r}{1+r} \delta_{m, n+1}+ \frac{1}{1+r} \delta_{m, n-1} -  \delta_{m,n}.
\end{equation}
Since we are interested in the fixation-time distribution, we condition on fixation occurring. As discussed in Section~\ref{visitStatistics} (see also Supplemental Material, Section I \cite{SM}), the conditioned transition matrix $\tilde{\Omega}_\text{tr} = S \, \Omega_\text{tr} \, S^{-1}$, where $S$ is diagonal with $S_{mm} = 1-1/r^m$. Combining these results, we have that \begin{equation}
    \tilde{\Omega}_\text{tr} = S \, \Omega_\text{RW} \, S^{-1} D,
\end{equation}
where we have used the fact that both $D$ and $S$ are diagonal matrices, and therefore commute.

We found in Section~\ref{visitStatistics} that the moments of the fixation time $T$ can be expressed as, 
\begin{equation}\label{fixationMomentsApp}
    \mu_n \coloneqq E[T^n] = (-1)^n n! \, \mathbf{1} \, \tilde{\Omega}_\text{tr}^{-n} \,  \mathbf{p}_\text{tr}(0),
\end{equation}
where $\mathbf{1}$ is a row vector of ones and $\mathbf{p}_\text{tr}(0)$ is the initial state of the system, with $[p_\text{tr}(0)]_m = \delta_{m,m_0}$ for $m_0$ initial mutants. To compute these moments, we need the inverse $\tilde{\Omega}_\text{tr}^{-1} = D^{-1} S \, \Omega_{RW}^{-1} S^{-1}$. Since $\Omega_{RW}$ is a tridiagonal Toeplitz matrix, its inverse has a well-known form \cite{daFonseca2001explicit}:
\begin{equation}
    (-\Omega_{RW})^{-1}_{ij} = \begin{cases}
                \displaystyle \frac{(r+1)(r^i-1)(r^N-r^j)}{r^j \, (r-1)(r^N-1)} \quad \text{if } \, i \leq j, \\[.75em]
                \displaystyle \frac{(r+1)(r^j-1)(r^N-r^i)}{r^j \, (r-1)(r^N-1)} \quad \text{if } \, i > j. \\
             \end{cases}
\end{equation}
Hence the matrix $V = -S\, \Omega_\text{RW}^{-1} S^{-1}$ has elements
\begin{equation}\label{meanVisitsApp}
   V_{ij} = \begin{cases}
                \displaystyle \frac{(r+1)(r^i-1)^2(r^N-r^j)}{r^i \, (r-1)(r^j-1)(r^N-1)} \quad \text{if } \, i \leq j, \\[.75em]
                \displaystyle \frac{(r+1)(r^i-1)(r^N-r^i)}{r^i \, (r-1)(r^N-1)} \quad \text{if } \, i > j. \\
             \end{cases}
\end{equation}
The matrix $V$, sometimes called the fundamental matrix, encodes the visit statistics of the conditioned random walk: $V_{ij}$ is the mean number of visits to state $i$ from state $j$ before hitting the absorbing state $N$ \cite{kemeny1983finite}. The Moran process has the same visit statistics, but on average spends a different amount of time, designated by $(b_i + d_i)^{-1}$, waiting in each state. 

While one could now compute the moments $\mu_n$ in Eq.~(\ref{fixationMomentsApp}) directly, we find that the cumulants yield nicer expressions. Furthermore, the normal and Gumbel fixation-time distributions, predicted by our simulations and approximate calculations, are more simply described in terms of their cumulants. The non-standardized cumulants $\kappa'_n$ are linear combinations involving products of moments whose orders sum to $n$. Thus each term in the cumulants has $n$ powers of $D$ producing $n$ factors of $(b_i+d_i)^{-1}$ with a weight designated by the visit statistics. With this observation, it is clear the standardized cumulants $\kappa_n = \kappa'_n/(\kappa'_2)^{n/2}$ have the form given in Eq.~(\ref{exactCumulants}),
\begin{equation}\label{exactCumulantsApp}
    \kappa_n(N) = \frac{\displaystyle \sum_{i_1,i_2, \dots,i_n=1}^{N-1} \frac{w_{i_1i_2 \cdots i_n}^n(r,N | m_0)}{(b_{i_1}+d_{i_1})(b_{i_2}+d_{i_2})\cdots(b_{i_n}+d_{i_n})}}{ \displaystyle \left(\sum_{i,j=1}^{N-1} \frac{w_{ij}^2(r,N |m_0)}{(b_i+d_i)(b_j+d_j)} \right)^{n/2}},
\end{equation}
where $w_{i_1i_2 \cdots i_n}^n(r,N|m_0)$ are the weighting factors coming entirely from the visit statistics of a biased random walk (starting from $m_0$ initial mutants). As in the main text, we take the initial state to be a single mutant $m_0=1$, and will suppress the dependence of the weighting factors on initial condition, writing $w_{i_1i_2 \cdots i_n}^n(r,N)$ instead. Generalizations to other cases are straightforward and are discussed briefly below. 

We emphasize that even without explicit knowledge of the factors $w_{i_1i_2 \cdots i_n}^n(r,N)$, this formulation can be extremely useful. For instance when $b_i + d_i$ is constant, these are just the cumulants for the (possibly biased) random walk, which were computed in Section~\ref{1D_lattice} to approximate the Moran process on the 1D lattice. In particular, the sums over weighting factors obtained from setting $b_i + d_i = 1$ in Eq.~(\ref{exactCumulantsApp}) have leading asymptotic form given by Eq.~(\ref{integralApprox}). This fact can be used to bound the cumulants even when $b_i + d_i \neq 1$, which in some cases is sufficient to determine the leading asymptotic behavior. When this is not possible, the weighting factors must be computed explicitly. We now turn our focus to derviing $w_{ij}^2(r,N)$ and $w_{ijk}^3(r,N)$. 

We can compute the weighting factors by writing out the matrix multiplication of $\tilde{\Omega}_\text{tr}^{-1}$. First note that
\begin{equation}
    [-\tilde{\Omega}_\text{tr}^{-1}]_{ij} = \frac{V_{ij}}{b_i + d_i}.
\end{equation}
Then the first three moments of the fixation time are,
\begin{equation}
\begin{split}
    \mu_1 &= \sum_{i=1}^{N} \frac{V_{i1}}{b_i+d_i}, \\
    \mu_2 &= 2 \sum_{i,j=1}^{N} \frac{V_{ij}V_{j1}}{(b_i+d_i)(b_j+d_j)}, \\
    \mu_3 &= 6 \sum_{i,j,k=1}^{N} \frac{V_{ij}V_{jk}V_{k1}}{(b_i+d_i)(b_j+d_j)(b_k+d_k)}.
\end{split}
\end{equation}
The corresponding non-standardized cumulants are given by the usual formulas, $\kappa'_2 = \mu_2-\mu_1^2$ and $\kappa'_3 = \mu_3 - 3 \mu_2 \mu_1 + 2 \mu_1^3$. In terms of the visit numbers the non-standardized cumulants become  
\begin{equation}\label{cumulantsVisitsApp}
\begin{split}
    \kappa'_2 &= \sum_{i,j=1}^N \frac{2\, V_{ij} V_{j1} - V_{i1} V_{j1}}{(b_i +d_i)(b_j+d_j)},\\
    \kappa'_3 &= \sum_{i,j=1}^N \frac{6\, V_{ij} V_{jk} V_{k1} - 6\, V_{ij}V_{j1} V_{k1} + 2\, V_{i1}V_{j1}V_{k1}}{(b_i+d_i)(b_j+d_j)(b_k+d_k)}.
\end{split}
\end{equation}
From here we can read off the weighting factors accordingly. For convenience, we can choose $w_{ij}^2(r,N)$ and $w_{ijk}^3(r,N)$ to be symmetric by averaging the numerators in Eq.~(\ref{cumulantsVisitsApp}) over the permutations of the indices. Then,
\begin{widetext}
\begin{equation}\label{weightsVisitsApp}
\begin{split}
    w_{ij}^2(r,N) &= \frac{1}{2} \sum_{\sigma \in \Pi_2} 2\, V_{\sigma(1) \sigma(2)} V_{\sigma(2)1} - V_{\sigma(1)1} V_{\sigma(2)1},\\
    w_{ijk}^3(r,N) &= \frac{1}{6}\sum_{\sigma \in \Pi_3} 6\, V_{\sigma(1)\sigma(2)} V_{\sigma(2)\sigma(3)} V_{\sigma(3)1} - 6\, V_{\sigma(1)\sigma(2)}V_{\sigma(2)1} V_{\sigma(3)1} + 2\, V_{\sigma(1)1}V_{\sigma(2)1}V_{\sigma(3)1},
\end{split}
\end{equation}
where $\Pi_2$ is the set of permutations of $\{i,j\}$ and $\Pi_3$ are the permutations of $\{i,j,k\}$. We note that these expressions also hold for general initial condition by replacing the subscript $1$ with $m_0$. Plugging Eq.~(\ref{meanVisitsApp}) into this expression for $w_{ij}^2$ we obtain, after some algebra,
\begin{equation}\label{weights2explicitApp}
    w_{ij}^2(r, N) = \frac{(r+1)^2 (r^j-1)^2(r^N-r^i)^2}{r^{i+j} \, (r-1)^2(r^N-1)^2},
\end{equation}
for $i\geq j$. Since we have constructed $w_{ij}^2(r,N)$ to be symmetric, when $j>i$ the formula is identical with $i$ and $j$ exchanged. Similarly, using Eq.~(\ref{meanVisitsApp}) together with the expression for $w_{ijk}^3$ in  Eq.~(\ref{weightsVisitsApp}) leads to
\begin{equation}\label{weights3explicitApp}
        w_{ijk}^3 (r,N) = 2\frac{(r+1)^3(r^k-1)^2(r^j-1)(r^N-r^i)^2(r^N-r^j)}{r^{i+j+k} \, (r-1)^3(r^N-1)^3},
\end{equation}
\end{widetext}
for $i\geq j \geq k$. Again, the formula for different orderings of the indices $i,j,k$ is the same with the indices permuted appropriately, so that $w_{ijk}^3$ is perfectly symmetric. 

This completes the derivation of the visit statistics expression for the fixation-time cumulants. Together, Eqs.~(\ref{exactCumulantsApp}), (\ref{weights2explicitApp}) and (\ref{weights3explicitApp}) give a closed form expression for the fixation-time skew which is manageable for the purpose of asymptotic approximations. The diagonal terms in the higher-order weighting factors are also particularly simple, $w_{ii \cdots i}^n(r,N) = (n-1)! V_{ii}^n$. While we will not explicitly compute them, the off diagonal weights $w_{i_1i_2 \cdots i_n}^n(r,N)$ can be found by a straightforward generalization of the above procedure. Example applications of this approach are given in Supplemental Material, Sections III and IV \cite{SM}, where we show that all cumulants of the fixation time vanish for the Moran process on the 1D lattice and compute the asymptotic skew for the Moran process on the complete graph.

\bibliography{references}

\end{document}


\title{Supplemental Material for ``Fitness dependence of the fixation-time distribution for evolutionary dynamics on graphs"}

\author{David Hathcock}
\affiliation{Department of Physics,  Cornell University, Ithaca, New York 14853, USA}
\author{Steven H. Strogatz}
\affiliation{Department of Mathematics,  Cornell University, Ithaca, New York 14853, USA}

\date{\today}

\maketitle

\onecolumngrid

This Supplemental Material is devoted to providing rigorous mathematical derivations supporting the results obtained in the main text. We start with the general theory for birth-death Markov chains. In Section~\ref{conditionedMarkovChain} we show how to condition the transition probabilities on fixation, producing a Markov chain with identical statistics that is guaranteed to reach fixation. This result is used both for the visit statistics approach (formulated in the main text Appendix, which gives an exact series expression for the fixation-time cumulants), and to derive a numerically efficient recurrence relation for calculating the fixation-time skew (Section~\ref{reccurenceRelation}). In Sections~\ref{1Dlattice} and \ref{completeGraph} we apply these results to the Moran process on the one-dimensional (1D) lattice and complete graph respectively. For the 1D lattice, we compute the asymptotic form of the fixation-time cumulants for neutral fitness and prove the cumulants vanish for non-neutral fitness. For the complete graph, we show the fixation-time skew under non-neutral fitness corresponds to that of a weighted convolution of Gumbel distributions and derive the $\alpha\rightarrow0$ limit of the truncated fixation-time skew in the Moran process with neutral fitness. Finally, in Section~\ref{twoFitness} we give further details regarding the extension of our results to the two-fitness Moran process.

\section{Birth-death Markov chain conditioned on fixation}\label{conditionedMarkovChain}

For both the numerical recurrence relation and the visit statistics approach, it is useful to consider the birth-death Markov chain conditioned on hitting $N$, which has an identical fixation-time distribution to the unconditioned process. This Markov chain has new conditioned transition probabilities denoted $\tilde{b}_m$ and $\tilde{d}_m$. If $X_t$ is the state of the system at time $t$, then $\tilde{b}_m \coloneqq \mathcal{P}(X_t = m \rightarrow X_{t+1} = m+1 | X_\infty = N)$ and $\tilde{d}_m$ is defined similarly. Applying the laws of conditional probability, we find that
\begin{equation}\label{conditionedB}
\begin{split}
\tilde{b}_m &= \frac{\mathcal{P}(X_{t+1} = m + 1 \text{ AND } X_t = m \text{ AND }  X_\infty = N)}{\mathcal{P}(X_t = m \text{ AND }  X_\infty = N)}\\
& =\frac{\mathcal{P}(X_\infty = N | X_{t} = m + 1) }{\mathcal{P}(X_\infty = N | X_t = m)} \mathcal{P}(X_{t+1} = m + 1 | X_t = m)\\
& =\frac{\mathcal{P}(X_\infty = N | X_{t} = m + 1) }{\mathcal{P}(X_\infty = N | X_t = m)} b_m,
\end{split}
\end{equation}
where $b_m$ is the transition rate in the original Markov chain. Following the same procedure, we find the backward transition probabilities are related by
\begin{equation}\label{conditionedD}
\tilde{d}_m =\frac{\mathcal{P}(X_\infty = N | X_{t} = m -1) }{\mathcal{P}(X_\infty = N | X_t = m)} d_m.
\end{equation}
The conditioned Markov chain has a few nice properties. First, the fixation probability in the conditioned system is one, by construction. This is particularly helpful for accelerating simulations of the Moran process. Conditioning the transition probabilities also accounts for the normalization of the fixation-time distribution. Furthermore, this operation only changes the relative probability of adding versus subtracting a mutant. The probability that the system leaves a given state is unchanged:
\begin{equation}
\begin{split}
\tilde{b}_m +\tilde{d}_m &= \frac{\mathcal{P}(X_{t+1} = m + 1 \text{ AND } X_t = m \text{ AND }  X_\infty = N) + \mathcal{P}(X_{t+1} = m - 1 \text{ AND } X_t = m \text{ AND }  X_\infty = N)}{\mathcal{P}(X_t = m \text{ AND }  X_\infty = N)}\\
& =1 - \frac{\mathcal{P}(X_{t+1} = m \text{ AND } X_t = m \text{ AND }  X_\infty = N)}{\mathcal{P}(X_t = m \text{ AND }  X_\infty = N)} \\
& = 1 - \mathcal{P}(X_{t+1} = m| X_t = m) \\
&= b_m + d_m.
\end{split}
\end{equation}
This invariance, along with Eqs.~(\ref{conditionedB}) and (\ref{conditionedD}), shows that conditioning the Markov chain is equivalent to a similarity transformation on the transient transition matrix with a diagonal change of basis:
\begin{equation}\label{similarityGeneral}
    \tilde{\Omega}_\text{tr} = S \, \Omega_\text{tr} \, S^{-1} \quad \quad S_{mn} = \mathcal{P}(X_\infty = N|X_t = m) \delta_{m,n},
\end{equation}
where $\Omega_\text{tr}$ is the birth-death transition matrix with absorbing states removed as defined in the main text.

For the Moran Birth-death process considered in the main text, $b_m/d_m = r$. In this case, by setting up a linear recurrence it is easy to show that the probability of fixation, starting from $m$ mutants, is
\begin{equation}
    \mathcal{P}(X_\infty = N | X_t = m) = \frac{1-1/r^{m}}{1-1/r^{N}},
\end{equation}
so that
\begin{equation}
    \tilde{b}_m = \frac{r^{m+1}-1}{r^{m+1}-r} b_m, \quad \quad \tilde{d}_m = \frac{r^m-r}{r^m-1} d_m.
\end{equation}
Note that we can scale the similarity matrix $S$ by an overall constant, so it is convenient to choose $S_{mn} = (1-1/r^m)\delta_{m,n}$. For the two-fitness Moran Birth-Death model discussed in the main text fixation probabilities derived by Kaveh et al. \cite{kaveh2015duality} can be used together with Eq.~(\ref{similarityGeneral}) to condition the Markov chain on fixation.

\section{Recurrence relation for fixation-time skew}\label{reccurenceRelation}
With the conditioned transition probabilities derived in Section~\ref{conditionedMarkovChain}, there is a reflecting boundary at $m=1$, which lets us set up a recurrence relation for the fixation-time moments. This derivation follows the method described by Keilson in Ref.~\cite{keilson1965review}. Let $S_m(t)$ be the first passage time densities from state $m$ to state $m+1$. Clearly, $S_1(t)$ has an exponential distribution,
\begin{equation}
    S_1(t) = \tilde{b}_1 e^{-\tilde{b}_1 t}.
\end{equation}
From $m>1$, the state $m+1$ can be reached either directly, with exponentially distributed times, or indirectly by first stepping backwards to $m-1$, returning to $m$, and then reaching $m+1$ at a latter time. Thus, the densities $S_m(t)$ satisfy
\begin{equation}
    S_m(t) = \tilde{b}_m e^{-(\tilde{b}_m + \tilde{d}_m) t} + \tilde{d}_m e^{-(\tilde{b}_m + \tilde{d}_m) t} * S_{m-1}(t) * S_{m}(t),
\end{equation}
where the symbol $*$ denotes a convolution. This equation can be solved by Fourier transform to obtain 
\begin{equation}\label{recRelation}
    S_m(\omega) = \frac{\tilde{b}_m}{\tilde{b}_m + \tilde{d}_m - \tilde{d}_m S_{m-1}(\omega) - i \omega }.
\end{equation}

We can compute a recurrence relation for the moments of the first passage time densities $S_m(t)$ by differentiating Eq.~(\ref{recRelation}). Let $\mu_m$, $\nu_m$ and $\gamma_m$ to be the first, second, and third moments of $S_m(t)$ respectively. Using the relations $\nu_m = -i S'(\omega=0)$, $\xi_m = (-i)^2 S''(\omega=0)$, and $\zeta_m = (-i)^3 S'''(\omega=0)$, we find that
\begin{equation}\label{recRelations}
\begin{split}
    \nu_m &= \tilde{b}_m^{-1}(1+\tilde{d}_m \nu_{m-1}), \\
    \xi_m &= \tilde{b}_m^{-2}[\tilde{b}_m \tilde{d}_m \xi_{m-1} +  2 (1+\tilde{d}_m \nu_{m-1})^2], \\
    \zeta_m &=\tilde{b}_m^{-3}[\tilde{b}_m^2 \tilde{d}_m \zeta_{m-1} + 6 \tilde{b}_m \tilde{d}_m \xi_{m-1} (1+\tilde{d}_m \nu_{m-1}) +  6 (1+\tilde{d}_m \nu_{m-1})^3],
\end{split}
\end{equation}
with boundary conditions $\nu_0 = \xi_0 = \zeta_0 = 0$. The recurrence relations in Eq.~(\ref{recRelations}) give the moments of incremental first passage time distributions $S_m(t)$. The total fixation time, $T$ is the sum of these incremental first passage times. Thus, the cumulants of $T$ are the sum of the cumulants of the incremental times and the skew of $T$ can be expressed as, 
\begin{equation}\label{recurenceSkew}
    \kappa_3(N) =  \left(\sum_{m=1}^{N-1} \zeta_m - 3 \xi_m \nu_m + 2 \nu_m^3 \right)\Bigg/ \left(\displaystyle \sum_{m=1}^{N-1}\xi_m-\nu_m^2 \right)^{3/2}.
\end{equation}
Numerical computation of for $\kappa_3(N)$ requires calculating the $3N$ moments and carrying out the two sums in Eq.~(\ref{recurenceSkew}). By bottom-up tabulation of the incremental moments, this procedure can be completed in $\mathcal{O}(N)$ time, asymptotically faster than the eigenvalue decomposition and the exact series solution from visit statistics.

\section{Asymptotic Analysis for the 1D Lattice }\label{1Dlattice}

\subsection{Neutral Fitness}

As in the main text, we begin with the neutral fitness Moran process on a 1D lattice with periodic boundary conditions. In this case, the eigenvalues of the transition matrix describing the system are,
\begin{equation}\label{1DneutralEigs}
    \lambda_m = \frac{2}{N} - \frac{2}{N}\cos \bigg (\frac{m \pi}{N}\bigg), \quad m = 1, 2, \dots, N-1.
\end{equation}
From the eigen-decomposition of the Markov birth-death process described in the main text, the standardized fixation-time cumulants are given by 
\begin{equation}\label{fixation_cumulants}
\kappa_n(N) = (n-1)! \left( \sum_{m =1}^{N-1} \frac{1}{\lambda_m^n} \right) \Bigg/  \left( \sum_{m =1}^{N-1} \frac{1}{\lambda_m^2} \right)^{n/2}.
\end{equation}
Note that the constant factor $2/N$ cancels in Eq.~(\ref{fixation_cumulants}), so we may equivalently consider rescaled eigenvalues $\lambda_m = 1-\cos(m\pi/N)$. To derive the asymptotic cumulants, we compute the leading asymptotic behavior of sums
\begin{equation}
    S_n = \sum_{m=1}^{N-1}\frac{1}{[1 - \cos(m\pi/N)]^n}.
\end{equation}
The function $ (1-\cos x)^{-n}$ can be expanded as a Laurent series $\sum_{k=0}^\infty c_k(n) x^{2(k-n)}$, which is absolutely convergent for $x\neq 0$ in the interval $(-2\pi, 2\pi)$. So the sum $S_n$ can then be expressed as 
\begin{equation}\label{sumAsymptotics}
\begin{split}
S_n &= \sum_{m=1}^{N-1} \sum_{k=0}^\infty c_k(n) \bigg(\frac{\pi m}{N} \bigg)^{2(k-n)}\\
&= \sum_{k=0}^\infty c_k(n) (N/\pi)^{2(n-k)} H_{N-1,2(n-k)}\\
&= \frac{c_0(n) \zeta(2n)}{\pi^{2n}}N^{2n} + \mathcal{O}(N^{2(n-1)})
\end{split}
\end{equation}
where $H_{N,q} = \sum_{m=1}^N m^{-q}$ is the generalized harmonic number and in the last line we used the asymptotic approximation
\begin{equation}
H_{N,2q} = 
    \begin{cases} 
        \displaystyle \zeta(2q)  + \mathcal{O}(N^{1-2q}) & q>0, \\[.75em]
        \displaystyle \frac{N^{1-2q}}{2q+1} + \mathcal{O}(N^{-2q}) & q\leq 0. \\
    \end{cases}
\end{equation}
It is easy to check that $c_0(n) = 2^n$. Now the cumulants are $\kappa_n(N) = (n-1)! S_n/S_2^{n/2}$, which for $N\rightarrow \infty$ are
\begin{equation}
\begin{split}
    \kappa_n &= (n-1)! \left(\frac{2^n \zeta(2n)}{\pi^{2n}} \right) \Bigg/ \left(\frac{2^2 \zeta(4)}{\pi^{4}} \right)^{n/2} \\
    &= (n-1)! \, \frac{\zeta(2n)}{\zeta(4)^{n/2}},
\end{split}
\end{equation}
as reported in the main text.

\subsection{Non-neutral fitness}

For non-neutral fitness, we showed in the main text that in the random walk approximation the fixation-time distribution is asymptotically normal. Here we use the visit statistics approach to prove this holds even when the variation in time spent in each state is accounted for. From the visit statistics formulation, the standardized cumulants of the fixation time (starting from a single initial mutant) can be written as,
\begin{equation}\label{exactCumulants}
    \kappa_n(N) = \left(\sum_{i_1,i_2, \dots,i_n=1}^{N-1} \frac{w_{i_1i_2 \cdots i_n}^n(r,N)}{(b_{i_1}+d_{i_1})(b_{i_2}+d_{i_2})\cdots(b_{i_n}+d_{i_n})} \right) \Bigg/ \left(\sum_{i,j=1}^{N-1} \frac{w_{ij}^2(r,N)}{(b_i+d_i)(b_j+d_j)} \right)^{n/2},
\end{equation}
where $w_{i_1i_2 \cdots i_n}^n(r,N)$ are the weighting factors that depend on the visit statistics of a biased random walk. To prove the fixation-time distribution is normal, we derive bounds on the sums
\begin{equation}
    S_n = \sum_{i_1,i_2, \dots,i_n=1}^{N-1} \frac{w_{i_1i_2 \cdots i_n}^n(r,N)}{(b_{i_1}+d_{i_1})(b_{i_2}+d_{i_2})\cdots(b_{i_n}+d_{i_n})}
\end{equation}
that appear in Eq.~(\ref{exactCumulants}) and show that $\kappa_n(N) \rightarrow 0$ as $N\rightarrow \infty$. First, note that $w_{i_1 i_2 \cdots i_n}^n(r,N) = (n-1)! \, V_{ii}^n \geq 1$ if $i_1=i_2=\dots =i_n \equiv i$. Furthermore, we claim that $w_{i_1 i_2 \cdots i_n}^n(r,N) \geq 0$ for all $i_1, i_2,\dots,i_n$. If this were not the case, one could construct a birth-death process with negative fixation-time cumulants by choosing $b_i+d_i$ appropriately. But we know the fixation-time cumulants are positive from the eigen-decomposition described in the main text. With these observations, we can bound $S_n$ from below by the sum over unweighted diagonal elements.  Similarly, the sums are bounded from above by the maximum value of $(b_i + d_i)^{-n}$ times the sum over the weighting factors. Putting these together, we obtain
\begin{equation}\label{generalBounds}
    \sum_{i=1}^{N-1} \frac{1}{(b_i+d_i)^n} \leq S_n \leq \left(\max_{1\leq i <N} \frac{1}{b_i + d_i} \right)^n \times \sum_{i_1,i_2, \dots,i_n=1}^{N-1} w_{i_1i_2 \cdots i_n}^n(r,N).
\end{equation}

The Moran process on the 1D lattice has transition probabilities $b_i + d_i = (1+r)/(r m + N-m)$. Then, as $N \rightarrow \infty$, the lower bound is
\begin{equation}\label{lowerBound}
    \sum_{i=1}^{N-1} \frac{1}{(b_i+d_i)^n} = \frac{1}{(r+1)^n}\sum_{m=1}^{N-1}{(r m + N - m)^n} = \frac{1+r+r^2+ \cdots + r^n}{(n+1)(1+r)^n}N^{n+1} + \mathcal{O}(N^n). 
\end{equation}
For the upper bound, first note that
\begin{equation}\label{upperBound1}
  \left(\max_{1\leq i <N} \frac{1}{b_i + d_i} \right)^n = \left[r(N-1)+1\right]^n = r^n N^n + \mathcal{O}(N^{n-1}).
\end{equation}
The sums over the weighting factors give the (non-standardized) fixation-time cumulants corresponding to a process with $b_i+d_i = 1$ and uniform bias $r$. This is exactly the biased random walk model used to approximate the Moran process in the main text. It follows that as $N\rightarrow \infty$,
\begin{equation}
    \sum_{i_1,i_2, \dots,i_n=1}^{N-1} w_{i_1i_2 \cdots i_n}^n(r,N) = (n-1)! \sum_{i=1}^{N-1} \left(\frac{1}{1-2\sqrt{r}/(r+1) \cos(m \pi/N)} \right)^n,
\end{equation}
where the denominators in the second sum are the eigenvalues of the transition matrix for the biased random walk, $\lambda_m = 1-2\sqrt{r}/(r+1) \cos(m \pi/N)$. As in the main text, we can estimate the leading asymptotics of this sum by converting to an integral,
\begin{equation}\label{upperBound2}
    \sum_{i_1,i_2, \dots,i_n=1}^{N-1} w_{i_1i_2 \cdots i_n}^n(r,N) = \frac{N}{\pi}  \int_0^\pi \frac{(n-1)! }{(1-2 \sqrt{r}/(1+r) \cos x)^n} \, \text{d}x + \mathcal{O}(1).
\end{equation}

Combining the results from Eqs.~(\ref{generalBounds})--(\ref{upperBound1}) and (\ref{upperBound2}) we arrive at
\begin{equation}\label{explicitBounds}
    \frac{1+r+r^2+ \cdots + r^n}{(n+1)(1+r)^n}N^{n+1} + \mathcal{O}(N^n) \leq S_n \leq \frac{N^{n+1}}{\pi}  \int_0^\pi \frac{(n-1)! }{(1-2 \sqrt{r}/(1+r) \cos x)^n} \, \text{d}x + \mathcal{O}(N^n).
\end{equation}
For each $n$, our upper and lower bounds have the same asymptotic scaling as a power of $N$, with different $r$-dependent coefficients. Using these results together in Eq.~(\ref{exactCumulants}), it follows that for $N \gg 1$, the cumulants to leading order are 
\begin{equation}
    \kappa_n(N) = C_n(r) \frac{1}{N^{(n-2)/2}} + \mathcal{O}(N^{-n/2}),
\end{equation}
where $C_n(r)$ is a fitness-dependent constant. Thus, indeed $\kappa_n(N) \rightarrow 0$ as $N\rightarrow \infty$. 

This result confirms the claim made in the main text. Even with heterogeneity in the time spent in each state, the skew and higher-order cumulants of the fixation time vanish asymptotically. Therefore, the Moran Birth-death process on the 1D lattice with non-neutral fitness $r>1$ has an asymptotically normal fixation-time distribution. The normal distribution is universal, independent of fitness level for this population structure.

\section{Asymptotic Analysis for the Complete Graph }\label{completeGraph}

\subsection{Non-neutral fitness}
In the main text we predicted that the asymptotic fixation-time distribution for the Moran Birth-death process on the complete graph is a convolution of two Gumbel distributions by applying our intuition from coupon collection. Furthermore, our calculation of the fixation-time cumulants in the large (but finite) fitness limit agrees with this prediction. Surprisingly, numerical calculations using the recurrence relation formulated above and direct simulations of the Moran process indicate that this result holds for all $r>1$. In this section we prove, using the visit statistics formulation, that the asymptotic skew of the fixation time for $r>1$ is identical to that of a convolution of Gumbel distributions. Based on our numerical evidence, we conjecture that an analogous calculation holds to all orders. The below calculation also shows why the coupon collection heuristic works: the asymptotically dominant terms come exclusively from the regions near fixation ($m=N-1$) and near the beginning of the process when a single mutant is introduced into the system ($m=1$).

As for the 1D lattice, we want to derive the asymptotic behavior of the sums 
\begin{equation}\label{completeSn}
    S_n = \sum_{i_1,i_2, \dots,i_n=1}^{N-1} \frac{w_{i_1i_2 \cdots i_n}^n(r,N)}{(b_{i_1}+d_{i_1})(b_{i_2}+d_{i_2})\cdots(b_{i_n}+d_{i_n})},
\end{equation}
where the transition probabilities $b_i$ and $d_i$ are those for the Moran process on the complete graph,
\begin{equation}
    b_i + d_i = \frac{(1+r)i(N-i)}{(N-1)(r i + N - i)}
\end{equation}
and the weights $w^2_{ij}(r,N)$ and $w^3_{ijk}(r,N)$ are respectively given by
\begin{equation}\label{weights2explicit}
    w_{ij}^2(r, N) = \frac{(r+1)^2 (r^j-1)^2(r^N-r^i)^2}{r^{i+j} \, (r-1)^2(r^N-1)^2}
\end{equation}
for $i>j$ and 
\begin{equation}
        w_{ijk}^3 (r,N) = 2\frac{(r+1)^3(r^k-1)^2(r^j-1)(r^N-r^i)^2(r^N-r^j)}{r^{i+j+k} \, (r-1)^3(r^N-1)^3},
\end{equation}
for $i>j>k$. The expressions for different orderings of indices are the same but with the indices permute appropriately so that $w^2_{ij}$ and $w^3_{ijk}$ are perfectly symmetric.

To start, consider the sums Eq.~(\ref{completeSn}), but with two indices $i_1$ and $i_2$ constrained to integers from $\alpha N$ to $(1-\alpha) N$ for $1/2>\alpha>0$. This sum may be written as 
\begin{equation}\label{twoIndexSums}
    S_n^{\alpha} = \sum_{i_1,i_2 = \alpha N}^{(1-\alpha) N} \sum_{i_3,\cdots,i_n = 1}^{N-1}\frac{w_{i_1i_2 \cdots i_n}^n(r,N)}{(b_{i_1}+d_{i_1})(b_{i_2}+d_{i_2})\cdots(b_{i_n}+d_{i_n})}.
\end{equation}
Now we may apply the upper bound in Eq.~(\ref{generalBounds}), but for the sums restricted to $\alpha N < i,j< (1-\alpha) N$, the maximum of $(b_i+d_i)^{-1}$ can also be restricted to this range,
\begin{equation}
\begin{split}
    S_n^\alpha &\leq \left(\max_{1< i< N} \frac{1}{b_i+d_i} \right)^{n-2} \times \left(\max_{\alpha N < i< (1-\alpha) N} \frac{1}{b_i+d_i} \right)^{2} \times \sum_{i_1,i_2,\cdots,i_n =1}^N w_{i_1i_2 \cdots i_n}^n(r,N)\\
    &= N^{n-1} \left\{ \left(\frac{r}{1+r}\right)^{n-2} \left(\frac{r(1-\alpha) + \alpha}{(1+r)(1-\alpha)\alpha} \right)^{2} \times \frac{1}{\pi}  \int_0^\pi \frac{(n-1)! }{(1-2 \sqrt{r}/(1+r) \cos x)^n} \, \text{d}x \right\} + \mathcal{O}(N^{n-2}).
\end{split}
\end{equation}
In the second line we used the integral approximation from Eq.~(\ref{upperBound2}) and evaluated the maximum of $(b_i+d_i)^{-1}$ over the indicated intervals. Since we constructed $w_{i_1i_2 \cdots i_n}^n(r,N)$ to be symmetric, this upper bound holds for any permutation of the indices in Eq.~(\ref{twoIndexSums}).

We now consider the same sums but with $1<i_1<\alpha N$ or $(1-\alpha)N < i_1 < N-1$,  
\begin{equation}\label{oneIndexSum1}
    S_n^{\alpha, 1} = \sum_{i_1 = 1}^{\alpha N} \sum_{i_2 = \alpha N}^{(1-\alpha) N} \sum_{i_3,\cdots,i_n = 1}^{N-1}\frac{w_{i_1i_2 \cdots i_n}^n(r,N)}{(b_{i_1}+d_{i_1})(b_{i_2}+d_{i_2})\cdots(b_{i_n}+d_{i_n})}.
\end{equation}
and 
\begin{equation}\label{oneIndexSum2}
    S_n^{\alpha, 2} = \sum_{i_1 = (1-\alpha)N}^{N-1} \sum_{i_2 = \alpha N}^{(1-\alpha) N} \sum_{i_3,\cdots,i_n = 1}^{N-1}\frac{w_{i_1i_2 \cdots i_n}^n(r,N)}{(b_{i_1}+d_{i_1})(b_{i_2}+d_{i_2})\cdots(b_{i_n}+d_{i_n})}.
\end{equation}
These sums can be estimated using the same upper bound, but without extending the sum on $w_{i_1i_2 \cdots i_n}^n(r,N)$ to the entire domain. Specifically, 
\begin{equation}\label{oneIndexApprox}
    S_n^{\alpha, 1} \leq N^{n-1} \left\{ \left(\frac{r}{1+r}\right)^{n-1} \left(\frac{r(1-\alpha) + \alpha}{(1+r)(1-\alpha)\alpha} \right) \right \} \times \sum_{i_1 = 1}^{\alpha N} \sum_{i_2 = \alpha N}^{(1-\alpha) N} \sum_{i_3,\cdots,i_n = 1}^{N-1} w_{i_1i_2 \cdots i_n}^n(r,N) + \mathcal{O}(N^{n-2}).
\end{equation}
Note that the weighting factors fall off exponentially away from the diagonal elements. This is because the visit numbers in the biased random walk become only very weakly correlated if the states are far away from each other. Thus, the sum in Eq.~(\ref{oneIndexApprox}) over terms away from the diagonal elements converges to a constant as $N\rightarrow \infty$. We have verified this explicitly for $w_{ij}^2(r,N)$ and $w_{ijk}^3(r,n)$. The series $S_n^{\alpha, 2}$ is similarly bounded, as are all sums of the form Eq.~(\ref{oneIndexSum1}) or (\ref{oneIndexSum2}) with the indices permuted.

The remaining terms in $S_n$ involve all indices in either $[1, \alpha N]$ or $[(1-\alpha) N, N-1]$. If not all indices are in the same interval, the weighting factors are exponentially small: the visit numbers near $m=1$ are uncorrelated with those near $m=N-1$. Thus each term in the sum is exponentially suppressed and doesn't contribute to $S_n$ asymptotically. With this observation only two parts of the sum remain: those with bounds $1\leq i_1, i_2 \dots i_n\leq \alpha N$ or $(1-\alpha) N\leq i_1, i_2 \dots i_n\leq N-1$. We call the sums with these bounds $S_n^{c1}$ and $S_n^{c2}$ respectively. As we will see below, the sums over these regions have leading order $\mathcal{O}(N^n)$. Since all the above terms are order $\mathcal{O}(N^{n-1})$ or smaller, the asymptotic behavior of the cumulants is entirely determined by these regions near the beginning and end of the process, i.e. the coupon collection regions. The fact that we can restrict the sums to this region allows us to make approximations that do not change the leading asymptotics, but make the sums easier to carry out. For instance, in $S_2^{c1}$, we can set $r^N-r^i \rightarrow r^{N}$ and $(N-i) \rightarrow N$, since the indices run only up to $\alpha N$. This gives 
\begin{equation}
\begin{split}
    S_2^{c1} &= \frac{N^2}{(r-1)^2} \left \{\sum_{i=1}^{\alpha N} \frac{(r^i-1)^2}{i^2 r^{2i} } + 2 \sum_{i=1}^{\alpha N} \sum_{j = 1}^{i-1} \frac{(r^j-1)^2}{i j \,  r^{i+j}} \right \} + \mathcal{O}(N) \\
    & = \frac{N^2 \zeta(2)}{(r-1)^2} + \mathcal{O}(N),
\end{split}
\end{equation}
for $N \gg 1$. A similar calculation shows $S_2^{c2} = r^2 N^2 \zeta(2)/(r-1)^2$. For the third order sums, we find
\begin{equation}
\begin{split}
    S_3^{c1} &= 2\frac{N^3}{(r-1)^3} \Bigg \{  \sum_{i=1}^{\alpha N} \frac{(r^i-1)^3}{i^3 r^{3i} } + 3 \sum_{i=1}^{\alpha N} \sum_{j = 1}^{i-1} \frac{(r^j-1)^3}{i j^2 r^{i+2j}} \\
    & \qquad \qquad \qquad \quad + 3 \sum_{i=1}^{\alpha N} \sum_{j = 1}^{i-1} \frac{(r^i-1)(r^j-1)^2}{i^2 j r^{2i+j}} + 6 \sum_{i=1}^{\alpha N} \sum_{j = 1}^{i-1}\sum_{k = 1}^{j-1} \frac{(r^j-1)(r^k-1)^2}{i j k \, r^{i+j+k}} \Bigg \} + \mathcal{O}(N^2) \\
    & = 2 \frac{N^3 \zeta(3)}{(r-1)^3} + \mathcal{O}(N^2),
\end{split}
\end{equation}
for $N \gg 1$. Again the other sum, with indices near $N-1$, is identical up to a factor of $r^3$, $S_3^{c2} = 2 r^3 N^3 \zeta(3)/(r-1)^3$. Overall, we have that
\begin{equation}
    S_2 =  \frac{N^2 (1+r^2) \zeta(2)}{(r-1)^2} + \mathcal{O}(N) \quad \quad \text{and} \quad \quad S_3 = \frac{2  N^3(1+r^3)\zeta(3)}{(r-1)^3} + \mathcal{O}(N^2).
\end{equation}
The asymptotic skew is given by 
\begin{equation}
    \kappa_3 = \frac{2 (1+r^3) \zeta(3)}{(r-1)^3} \Bigg / \left(\frac{(1+r^2)\zeta(2)}{(r-1)^2} \right)^{3/2} = \frac{1+r^3}{(1+r^2)^{3/2}} \times \frac{2 \zeta(3)}{\zeta(2)^{3/2}},
\end{equation}
which is exactly the skew corresponding to the convolution of Gumbel distributions with relative weighting given by the fitness, $G+rG'$. While evaluating the series to higher orders is increasingly difficult, our simulations and the large-fitness approximation suggest this result holds to all orders and that indeed, the asymptotic fixation-time distribution is a weighted convolution of Gumbel distributions.

\subsection{Neutral fitness with truncation}
As discussed in the main text, the neutral fitness Moran process on the complete graph has a fixation-time skew that depends on the level of truncation. That is, the time $T_\alpha$ it takes for the process to reach $\alpha N$ mutants, where $0 \leq \alpha \leq 1$, has a distribution whose skew depends on $\alpha$. Here we show that the $\alpha \rightarrow 0$ limit of the fixation-time skew equals $\sqrt{3}$. 

To start, we take the neutral fitness limit of the weighting factors to obtain 
\begin{equation}
    w_{ij}^2(1, \alpha N) = \frac{4 j^2 (\alpha N-i)^2}{\alpha^2 N^2}
\end{equation}
for $i\geq j$ and 
\begin{equation}
    w_{ijk}^3(1, \alpha N) = \frac{16 j k^2 (\alpha N-i)^2 (\alpha N - j)}{\alpha^3 N^3}
\end{equation}
for $i\geq j\geq k$, again with the expressions for other orderings obtained by permuting the indices accordingly. The neutral fitness Moran process on the complete graph has transition probabilities $b_i + d_i = 2(N i -i^2)/(N^2 -N)$. Since we are computing the truncated fixation-time skew, we use Eq.~(\ref{exactCumulants}), but cut the sums off at $\alpha N$. In this case, these sums are dominated by the off-diagonal terms, so that
\begin{equation}
\begin{split}
    S_2 = \sum_{i,j = 1}^{\alpha N} \frac{w^2_{ij}(1,\alpha N)}{(b_i+d_i)(b_j+d_j)}
    &= 2 \sum_{i = 1}^{\alpha N} \sum_{j = 1}^{i-1} \frac{j (\alpha N- i)^2 (N-1)^2}{\alpha^2 i(N-i)(N-j)} + \mathcal{O}(N^3)\\
    &= 2 \sum_{i = 1}^{\alpha N} \sum_{j = 1}^{i-1} \frac{j (\alpha N- i)^2}{\alpha^2 i} + \mathcal{O}(N^3)\\
    &= \frac{\alpha^2 N^4}{12} + \mathcal{O}(N^3),
\end{split}
\end{equation}
where in the second line we approximated $N-i$ and $N-j$ by $N$. This approximation is exact in the limit $\alpha \rightarrow 0$ since the upper limit on the sum, $\alpha N$, is much smaller than $N$. Using analogous approximations, we find 
\begin{equation}
\begin{split}
    S_3 = \sum_{i,j,k = 1}^{\alpha N} \frac{w^2_{ijk}(1,\alpha N)}{(b_i+d_i)(b_j+d_j)(b_k+d_k)} 
    &= 6 \sum_{i = 1}^{\alpha N} \sum_{j = 1}^{i-1} \sum_{k = 1}^{j-1}\frac{2 k (\alpha N- i)^2 (\alpha N-j) (N-1)^3}{\alpha^3 i(N-i)(N-j)(N-k)} + \mathcal{O}(N^5)\\
    &= 12 \sum_{i = 1}^{\alpha N} \sum_{j = 1}^{i-1} \sum_{k = 1}^{j-1} \frac{k (\alpha N- i)^2(\alpha N - j)}{\alpha^2 i} + \mathcal{O}(N^5) \\
    &= \frac{\alpha^3 N^6}{24} + \mathcal{O}(N^5).
\end{split}
\end{equation}
The asymptotic fixation-time skew as $\alpha \rightarrow 0$ is therefore 
\begin{equation}
    \kappa_3 = \frac{\alpha^3 N^6/24}{(\alpha^2 N^4/12)^{3/2}} = \sqrt{3},
\end{equation}
as claimed in the main text. This value agrees perfectly with our numerical calculations, which show the above approximation breaks down when $\alpha \approx 1/2$. Above this threshold, the random walk causes mixing between the two coupon collection regions, thereby lowering the overall skew of the fixation-time distribution toward the $\alpha = 1$ value of $\kappa_3 = 6 \sqrt{3} (10-\pi^2)/(\pi^2-9)^{3/2} \approx 1.6711$.

\section{Fixation-time distributions in the two-fitness Moran Process}\label{twoFitness}

As discussed in the main text, the two-fitness Birth-Death (BD) Moran process has the same family of fixation-time distributions as the Birth-death (Bd) process with only one fitness level. Here we provide further details leading to this conclusion. In particular, we give the transition probabilities for the two-fitness model and describe how the calculations from the main text generalize to this system. Here $r$ is the fitness level during the birth step, while $\tilde r$ is the fitness level during the death step in the Moran process.

\subsection{One-dimensional lattice}
 On the 1D lattice, the Moran process with fitness at both steps (birth and death), has new transition probabilities
 \begin{equation}\label{1Dtwofitness}
 b_m = \frac{r}{r m +N-m} \frac{\tilde{r}}{1+\tilde{r}}, \quad \quad d_m = \frac{1}{r \tilde{r}} b_m,
 \end{equation}
 for $1<m<N-1$. The probabilities are different when there is only one mutant or non-mutant ($m=1$ or $m = N-1$ respectively). In these cases the nodes on the population boundary don't have one mutant and one non-mutant as neighbors, as is the case for all other $m$. In the limit $N \gg 1$, however, changing these two probabilities does not affect the fixation-time distribution and we can use the probabilities given in Eq.~(\ref{1Dtwofitness}).

 The two-fitness Moran BD model on the 1D lattice differs from the previously considered Bd process in two ways. First, the transition probabilities have the same functional form as before, but are scaled by a factor $\tilde{r}(1+\tilde{r})^{-1}$. This factor determines the time-scale of the process but does not alter the shape of the fixation-time distribution because it drops out of the expressions for the cumulants, Eqs.~(\ref{fixation_cumulants}) and (\ref{exactCumulants}). Second, the ratio $b_m/d_m = r \tilde{r}$ shows that the process is still a random walk, but with new bias corresponding to an effective fitness level $r_\text{eff} = r \tilde{r}$. With these observations, when $r_\text{eff} \neq 1$, our preceding analysis applies and we predict normally distributed fixation times. If $r_\text{eff} = 1$, the random walk is unbiased, and we expect highly skewed fixation-time distributions.

\subsection{Complete Graph}

 On the complete graph, considering fitness during the replacement step leads to transition probabilities
 \begin{equation}\label{completeBirthtwo}
 b_m = \frac{r m}{r m + N-m} \cdot \frac{\tilde{r} (N-m)}{\tilde{r} (N-m) + m-1}, \quad \quad  d_m = \frac{N-m}{r m + N-m} \cdot \frac{m}{\tilde{r} (N-m-1) + m}.
 \end{equation}
 In this case, the ratio of transition probabilities is $m$-dependent, but $b_m/d_m \rightarrow r \tilde{r}$ as $N\rightarrow \infty$, again motivating the definition of the effective fitness level $r_\text{eff} = r \tilde{r}$. If we take the large (but not infinite) fitness limit $r_\text{eff} \gg 1$, so that the mutant population is monotonically increasing to good approximation, then the fixation time cumulants are again given by Eq.~(\ref{fixation_cumulants}) with $\lambda_m \rightarrow b_m+d_m$. As $N\rightarrow \infty$, the cumulants become
 \begin{equation}
 \kappa_n = \frac{1+r^n/\tilde{r}^n}{(1+r^2/\tilde{r}^2)^{n/2}} \cdot \frac{(n - 1)! \zeta(n)}{\zeta(2)^{n/2}},
 \end{equation}
 identical to the Moran Bd process on the complete graph, with $r \rightarrow r/\tilde{r}$. Numerical calculations (using the moment recurrence relation derived in Section~\ref{reccurenceRelation}) again indicate this expression for the cumulants holds for all $r$, not just in the large fitness limit. When $r_\text{eff} = 1$, we expect highly skewed fixation distributions arising from the unbiased random walk underlying the dynamics. This is indeed the case, though numerics indicate there is an entire family of distributions dependent on $r = 1/\tilde{r}$.

\bibliography{SI-references}